\documentclass[%
reprint,
superscriptaddress,
 aps,prx,
amsmath,amssymb]{revtex4-2}

\usepackage{graphicx}
\usepackage{dcolumn}
\usepackage{bm}

\usepackage{color}\usepackage{ulem}

\begin{document}

\title{Second-order microscopic nonlinear susceptibility in a centrosymmetric material: application to imaging valence electron motion. } 

\author{Chance Ornelas-Skarin}
\affiliation{Stanford PULSE Insitute, 2575 Sand Hill Rd, SLAC National Accelerator Labororatory, Menlo Park, CA 94025 USA}
\affiliation{Department of Electrical Engineering, Stanford University, Stanford, CA 94305 USA}
\author{Tatiana Bezriadina}
\affiliation{%
I. Institute for Theoretical Physics and Centre for Free-Electron Laser Science,
the University of Hamburg, Notkestr. 9, 22607 Hamburg, Germany\\}
\affiliation{The Hamburg Centre for Ultrafast Imaging (CUI),
Luruper Chaussee 149, 22607 Hamburg, Germany\\
}

\author{Matthias Fuchs}
\affiliation{Institute for Beam Physics and Technology (IBPT), Karlsruhe Institute of Technology (KIT), 76021 Karlsruhe, Germany}
\author{Shambhu Ghimire}
\affiliation{Stanford PULSE Insitute, 2575 Sand Hill Rd, SLAC National Accelerator Labororatory, Menlo Park, CA 94025 USA}
\author{J. B. Hastings}
\affiliation{Linear Coherent Light Source, SLAC National Accelerator Laboratory, Menlo Park, CA 94025, USA}
\affiliation{Stanford PULSE Insitute, 2575 Sand Hill Rd, SLAC National Accelerator Labororatory, Menlo Park, CA 94025 USA}
\author{Quynh L Nguyen}
\affiliation{Stanford PULSE Insitute, 2575 Sand Hill Rd, SLAC National Accelerator Labororatory, Menlo Park, CA 94025 USA}
\author{Gilberto de la Pe\~na}
\affiliation{Stanford PULSE Insitute, 2575 Sand Hill Rd, SLAC National Accelerator Labororatory, Menlo Park, CA 94025 USA}
\author{Takahiro Sato}
\affiliation{Linear Coherent Light Source, SLAC National Accelerator Laboratory, Menlo Park, CA 94025, USA}
\affiliation{Stanford PULSE Insitute, 2575 Sand Hill Rd, SLAC National Accelerator Labororatory, Menlo Park, CA 94025 USA}
\author{Sharon Shwartz}
\affiliation{Physics Department and Institute of Nanotechnology, Bar Ilan University, Ramat Gan 52900, Israel}
\author{Mariano Trigo}
\affiliation{Stanford PULSE Insitute, 2575 Sand Hill Rd, SLAC National Accelerator Labororatory, Menlo Park, CA 94025 USA}
\affiliation{Stanford Institute for Materials and Energy Sciences, SLAC National Accelerator Laboratory, Menlo Park, CA 94025, USA}
\author{Diling Zhu}
\affiliation{Linear Coherent Light Source, SLAC National Accelerator Laboratory, Menlo Park, CA 94025, USA}
\author{Daria Popova-Gorelova}
\affiliation{%
I. Institute for Theoretical Physics and Centre for Free-Electron Laser Science,
the University of Hamburg, Notkestr. 9, 22607 Hamburg, Germany\\}
\affiliation{The Hamburg Centre for Ultrafast Imaging (CUI),
Luruper Chaussee 149, 22607 Hamburg, Germany\\
}
\affiliation{Institute of Physics, Brandenburg University of Technology Cottbus-Senftenberg, Erich-Weinert-Stra\ss e 1, 03046 Cottbus, Germany
}%

\author{David A. Reis}
\affiliation{Stanford PULSE Insitute, 2575 Sand Hill Rd, SLAC National Accelerator Labororatory, Menlo Park, CA 94025 USA}

\date{\today}
\begin{abstract}

 We report measurements of phase-matched nonlinear x-ray and optical sum-frequency generation from single-crystal silicon using sub-resonant 0.95 eV laser pulses and 9.5 keV hard x-ray pulses from the LCLS free-electron laser. 
  The sum-frequency signal appears as energy and momentum sidebands to the elastic Bragg peak. It is proportional to the magnitude squared of the relevant temporal and spatial Fourier components of the optically induced microscopic charges/currents.
We measure the first- and second-order sideband to the 220 Bragg peak and find that the efficiency is maximized when the applied field is along the reciprocal lattice vector.  For an optical intensity of $\sim10^{12}  \text{W}/\text{cm}^2$, we measure peak efficiencies of $3\times 10^{-7}$ and $3\times 10^{-10}$ for the first and second-order sideband respectively (relative to the elastic Bragg peak). The first-order sideband is consistent with induced microscopic currents along the applied electric field (consistent with an isotropic response).  
 The second-order sideband depends nontrivially on the optical field orientation and is  consistent with an anisotropic response originating from induced charges along the bonds with C$_{3v}$ site symmetry.  The results agree well with first-principles Bloch-Floquet calculations.
 
\end{abstract}

\maketitle

\section{Introduction}

The microscopic, atomic-scale valence-electron density and its dielectric response are inextricably connected to both the static and dynamic properties of materials\cite{AschcroftMermin}.  It plays a fundamental role in bonding and the valence excitations that result in a material's macroscopic electrical, thermal, magnetic, and optical properties. The tetrahedrally- bonded group IV semiconductors are prototypical examples of the outsized role that the valence electron density, a small fraction of  the total electron density, can play in the structural stability of materials. In this case,  $sp^3$ hybridization leads to strong covalent bonding and stability against shear deformation\cite{Martin1968,Martin1969}.
Despite the important role of the valence charge density  
there is a dearth of methods for directly accessing its angstrom scale structure and excited-state dynamics.

Non-resonant x-ray scattering is a powerful tool for imaging atomic-scale structure, as the differential elastic scattering cross-section is proportional to the modulus square of the spatial Fourier transform of the electron density\cite{James}.  Conventional non-resonant x-ray scattering experiments provide this information from the entire electron density, valence and core. Because the valence density tends to be delocalized and make up only a small fraction of the total electron density, it can be difficult to separate from the more localized core density. In the case of the group-IV semiconductors,  the ground-state valence density can be obtained by detailed measurements of quasi-forbidden Bragg peaks 
after removing contributions from multiple scattering  and anharmonic thermal vibrations\cite{Batterman222,hastings1975high,Yangandcoppens,LuZungerDeuthsch}.  

X-ray and optical wavemixing (XOM) corresponds to the nonlinear interaction between single x-ray photons and $n$ optical photons mediated by  modulations of the valence charge density by both the x-ray and external optical excitation, or in the case of parametric down conversion through vacuum fluctuations. It was originally proposed by Freund and Levine\cite{Freund} and Eisenberger and McCall\cite{Eisenberger} around 1970, and has recently has received renewed attention for its potential to image  valence charge motion at optical frequencies with atomic-scale spatial resolution \cite{Glover,popova2018theory,Popova-GorelovaCommPhys24,Boemer,krebs,tamasaku2011visualizing,SchoriXPDtoOpt,SoferPDCUV}.  In the limit that the x-ray bandwidth is small compared to the optical period, phase-matched XOM in crystals appears as energy and momentum sidebands to ordinary Bragg peaks (corresponding to sum and difference frequency generation).  
The intensity of the sidebands, $I^{(n)}_G \propto |\rho^{(n)}_{\vec{G}}|^2$, where $\rho^{(n)}_{\vec{G}}$, are Fourier components of the electron density $\rho(\vec{r},t)$ corresponding to the $n^{\text{th}}$ harmonic of the optical frequency and $\vec{G}^{\text{th}}$ spatial Fourier component of the lattice. 
Thus, measurements of the amplitude and phase of the sidebands are necessary to reconstruct the motion of the atomic-scale optically induced valence electron density within the unit cell.  Since the optically induced charge density is a small fraction of the already small valence charge density, the sidebands are weak, and these measurements require high resolution analyzers to separate the sum (or difference) frequency signal from the elastic background.  

The first observation of XOM was reported in \cite{Glover} more than forty years after the original proposals.  In that experiment,  Glover \emph{et al.,} measured the first-order sum-frequency sideband between 8 keV x rays and 1.55 eV optical photons about the 111 Bragg peak from single crystal diamond. The results are consistent with first-principles calculations for the induced charge-density residing primarily along the tetrahedral bonds. Due to the lack of phase information, only the magnitude of the 111 spatial Fourier component of the charge density at the optical frequency could be extracted. Nonetheless, the results are a particularly notable measurement of the microscopic optical response, which is strongly affected by dielectric screening and local-field effects due to the high density and atomic-scale spatial inhomogeneities occurring on the scale of $\sim 10^{-4}$ of the optical wavelength\cite{Jackson}.   As we demonstrate here, higher-order wavemixing about a suitable Bragg peak  yields additional symmetry information that can help localize the valence density without the need of phasing.

In this article, we report x-ray and optical wave mixing in single-crystal silicon using monochromatic x-rays from a free-electron laser.  
We detect the first- and second-order sum-frequency sidebands about the 220 Bragg peak correspond to the nonlinear mixing of 9.5 keV x-ray photons with one and two 0.95 eV IR photons respectively. Silicon has the same centrosymmetric point group as diamond and thus a vanishing macroscopic second-order susceptibility in the dipole limit. Thus, the observation of a second-order sideband in a centrosymmetric material is an indication of broken inversion symmetry at the microscopic level.  

We find that the first-order sideband to the Si 220 behaves qualitatively like the first-order sideband to the diamond 111 reported by Glover \emph{et al.}. In contrast, the second-order sideband  corresponding to the 220 component of the local second-order susceptibility scales quadratically with the infrared pump intensity as expected for a perturbative nonlinear response. The IR- laser polarization dependence for the second-order sideband is inconsistent with the induced charges merely following the optical field, unlike what is seen in the first-order sideband reported here and in Ref.\cite{Glover}. This and the relatively high efficiency of the second-order sideband show that it is dominated by the local second-order dipole response, as opposed to the more general multipolar contribution.  

Our results are consistent with a second-order local nonlinear optical response originating from the optically induced motion of the interstitial charges with local inversion symmetry breaking and overall reduced symmetry compared to the atomic sites.     In particular, from a single measurement on (001) cut crystal, we constrain three of the four independent 220 spatial Fourier components of the local nonlinear optical susceptibility, $\chi^{(2)}_{11}(220) + \chi^{(2)}_{12}(220)  \approx \chi^{(2)}_{15}(220) $ (in Voigt notation).  
We further determine that $\chi^{(2)}_{11}(220) \approx 1.5\chi^{(2)}_{15}(220) $, and $\chi^{(2)}_{12}(220) \approx -0.6\chi^{(2)}_{15}(220)$ using a ($\bar{1}10$)-cut crystal. Our measurements agree well with our ab initio calculations for x-ray scattering from the induced charge density oscillating at twice the IR-laser frequency. We note however, that this second-harmonic component to the induced charge does not radiate, and thus can only be observed with an atomic scale probe such as x-ray optical mixing.
 Details of our formalism are given in Appendices \ref{a:chargedensity}--\ref{a:phasematching}.   In Section \ref{s:setup} we describe the experimental setup, followed by the procedure and results in Section \ref{s:results}. We conclude in Section \ref{s:conclusions} and give a brief outlook.

\section{Experimental Apparatus\label{s:setup}}

\begin{figure*}
\begin{center}
   \includegraphics[width=0.8\linewidth]{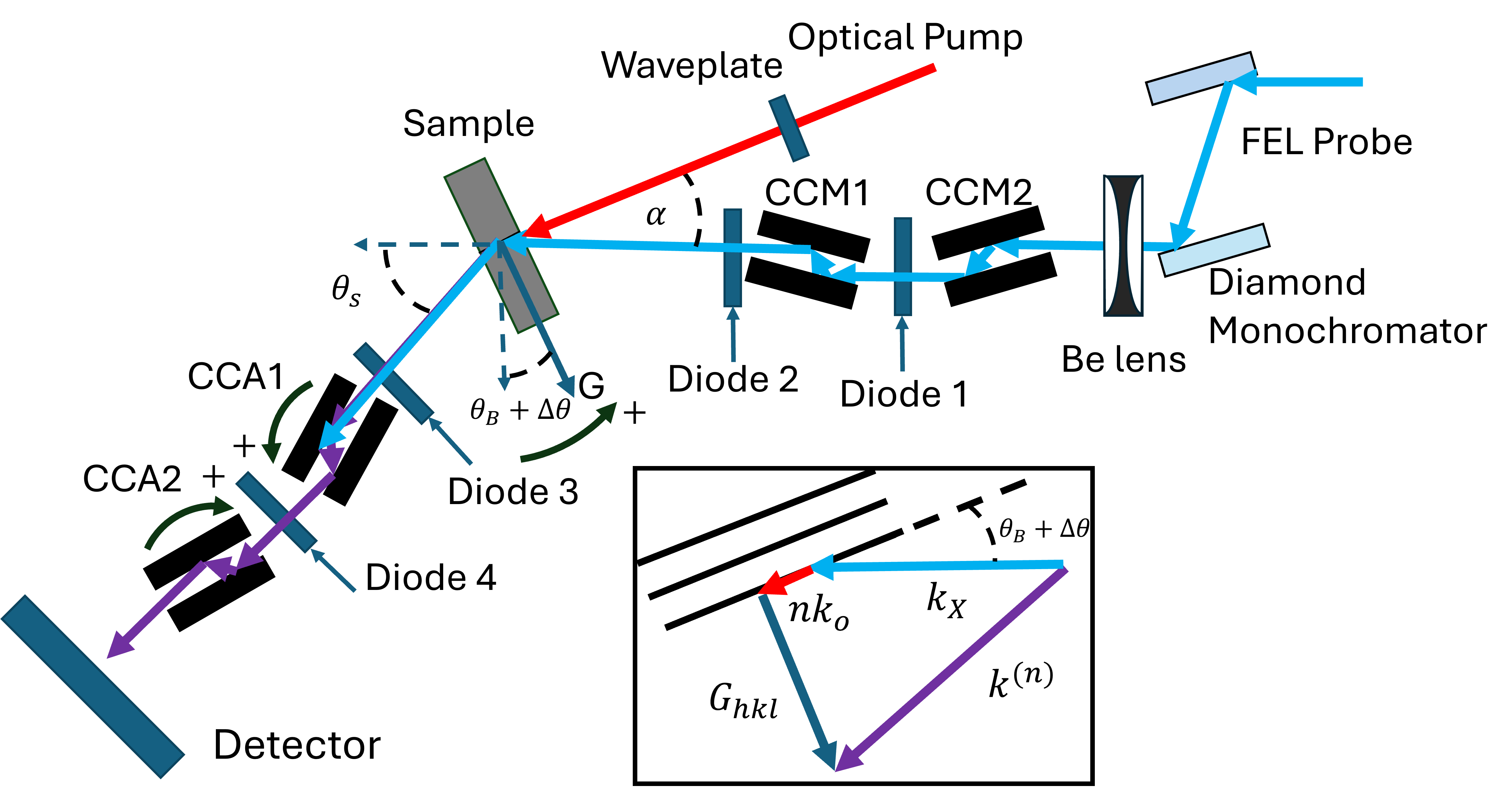}
   \caption{\label{fig:setup}  
Schematic of XOM experiment setup. From the right, the FEL hard x-ray beam at 9.5 keV (shown in light blue) is monochromatized by the XPP diamond 111 large offset monocrhomator, before being gently focused by compound Be lenses.  A secondary 311 Si channel cut monochromator reduces the bandwidth to $\sim$ 0.2 eV and re-collimates the beam in the horizontal scattering plane. The 0.95 eV optical pump passes through a half-wave plate for varying the polarization and a lens for better matching the spot size of the x-rays. It is coupled at near normal incidence to the sample surface at an angle $\alpha \approx \theta_{B}$. The sample is 40 $\mu$m thick (001)-cut Si. The two beams cross  at the sample location, and the sample is oriented to fulfill the phase-matching conditions for producing optical sidebands to the 220 Bragg peak (shown schematically for the first-order sideband in the inset on the bottom right). In the inset, $G_{hkl}$ is the reciprocal lattice vector ($G_{220}$ for our measurement), $k_x$ is the x ray wave-vector, $k^{(n)}$ is the wave-mixing wave-vector, and $nk_o$ is the corresponding wave-vector of n optical photons. Both the elastically scattered x rays from the tail of the Bragg peak (shown in blue after the sample) and the phase-matched XOM x rays (shown in purple after the sample) exit the crystal and enter the analyzer. The analyzer acceptance is narrow enough to allow the XOM photons onto the detector while rejecting the elastically scattered photons. Diodes 1--4 are placed along the beam path to allow real time monitoring of any drift between components. }
\end{center}
\end{figure*}

The measurements were performed using the X-ray Pump-Probe instrument at the LCLS hard x-ray free electron laser (FEL)\cite{XPP}. The FEL provided x-ray pulses at a repetition rate of 120 Hz with a central energy of 9.5 keV and a bandwidth of $\sim 1$ eV after the diamond 111 beamline monochromator (see Figure~\ref{fig:setup} ). The x rays are focused with a compound beryllium lens, to better match the optical spot size, and arrive at the sample position with an approximate spot size of $ \sim 20 \times 30 \mu\text{m}^2$ and pulse length of $\sim 30$ fs. The x rays are vertically polarized and the scattering plane for both the sample and crystal optics is in the horizontal plane. The sample is a 40 $\mu$m thick (001) cut, float zone single crystal silicon sample fabricated by Norcada. 

After the Be lenses the x-ray beam is further monochromatized and collimated by a custom 4 bounce 311 Si channel cut monochromator (CCM1 and CCM2) in a dispersive geometry with an estimated throughput of $\sim 5 \times 10^7$ photons/pulse (on average), a bandwidth of $\sim 0.2$ eV, and  divergence  $\leq 0.7$ mdeg (12 $\mu$rad) in the horizontal plane at the sample position.  The purpose of the monochromator is to create an incident x-ray beam that is sufficiently narrow in energy and divergence so that the sum-frequency signal can be well separated from the elastic background.

The sample is mounted using a manual Thorlabs RSP1 rotation mount to bring the 220 reciprocal lattice vector into the scattering plane (horizontal in the lab frame) for a symmetric Laue transmission scattering geometry. After the sample, we use a custom 4 bounce 311 Si channel cut analyzer (CCA1 and CCA2) matched to the monochromator (also in a dispersive geometry).  
 This provides an effective filter for accepting XOM photons while rejecting the elastically scattered background photons outside this acceptance. After the analyzer, we place a pixelated Jungfrau 1M detector that measures x-rays with single-photon sensitivity on a shot-by-shot basis. Additional lead shielding is used to minimize parasitic x-ray background from air-scattering and the various x-ray optics. All channel cut crystals (CCM1, CCM2, CCA1 and CCA2) and the sample are rotated with sub-millidegree resolution in the scattering plane using Kohzu RA10A-W01 rotation stages. 

 The sample is pumped with sub-bandgap 1300 nm (0.95 eV) $\sim 100$ fs, mJ-scale pulses provided by the XPP optical parametric amplifier. The relative arrival time between each optical and x-ray pulse is given on a shot-by-shot basis using an Ce:YAG-based arrival time monitor \cite{TimeTool}. The linearly polarized pump is focused with a 250 mm focal length lens to a measured spot size of $\sim 300\mu$m FWHM and peak intensity of $\sim 10^{12} \text{W}/\text{cm}^2$ at the sample position.  We choose to couple at near normal incidence so that reflection losses from the sample are approximately independent of the polarization of the laser. The linear polarization is rotated using a half-waveplate before the lens. 
 
 The x-ray flux is monitored on a shot-by-shot basis using photodiodes to measure scattered x rays from thin Kapton films.  The diodes (diode 1--4 in Figure~\ref{fig:setup}) are placed between CCM1 and CCM2, before the sample, after the sample, and between CCA1 and CCA2 to monitor the x-ray flux between components during the measurement. The optical laser is quasi-randomly mistimed with respect to the x rays in an overall ratio of 2 optical laser \emph{off} for every 3 optical laser \emph{on} shots, allowing for measurement of the background under otherwise similar conditions as the signal. In the figures below, the measured data are shown with the laser-on shots (laser-off shots) as blue circles (orange circles).  
 
\begin{figure*}
\begin{center}
   \includegraphics[width=0.75\linewidth]{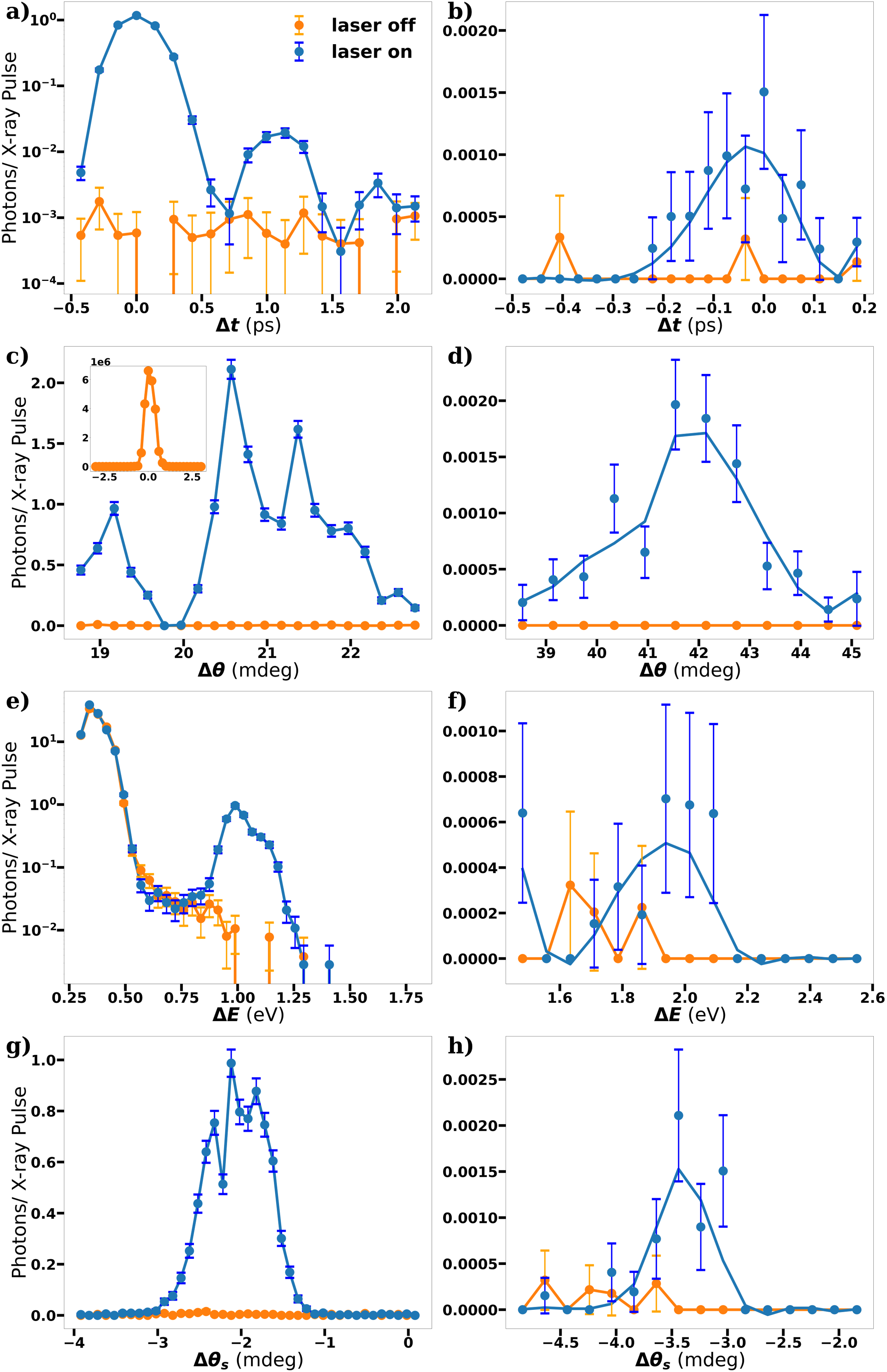}
    \caption{\label{fig:XOMdependance}
    Measurements of the first (left column) and second (right column) order XOM sidebands. Measurements with the optical laser on are shown in blue and the optical laser off are shown in orange. XOM dependence on relative delay $\Delta t$ between x-ray and optical pulse are shown for  a) first (on a logarithmic scale) and b) second order sidebands. XOM dependence on sample rotation $\Delta\theta$ in the scattering plane of the c) first and d) second order sidebands. The signal is maximized when the crystal is phase-matched, set to the peak of the sample rotation curve in c) or d), and temporal overlap is optimized, set to the peak of the relative delay curve in a) or b). Once the signal is maximized, the energy contents of the XOM beams measured with the analyzer are shown for the e) first order (on a logarithmic scale) and f) second order sideband. Note that the shoulder of the elastic Bragg peak is shown in e). Similarly, we show the dependence of the g) the first order sideband and h) the second order sideband on the scattering angle. }
\end{center}
\end{figure*}

\section{Procedure and Results\label{s:results}}

The x-ray optics and sample shown in  Figure ~\ref{fig:setup} are first aligned to the elastic 220 Bragg peak (at a Bragg angle $\theta_B=19.86^\circ$ for our x-ray energy). After aligning the Bragg peak the crystal is detuned according to the phase-matching condition (see Appendix  \ref{a:phasematching}) to isolate a particular side band and thus Fourier component of the induced charge density. The calculated sample detuning, $\Delta\theta$, for the first (second) order sideband of the Si 220 peak is 20.72 (41.44) mdeg. 
The scattering angle and energy of the analyzer are detuned by the calculated $\Delta\theta_s=1.9 (3.8) $ mdeg and $\Delta\theta_a=2.4 (4.8)$  mdeg from the nominal conditions ($2\theta_B\approx39.73^\circ$ and $\theta_a \approx 23.48^\circ$, respectively) for measuring the 220 elastic Bragg peak.  Here $\theta_a$ is the Bragg angle for the analyzers.   We rotate the channel cuts to set $\Delta\theta_s$ and $\Delta\theta_a$. Since the two analyzer crystals are not mounted on a common rotation about the sample, we compensate for the change in scattering angle (and analyzer energy) by rotating them individually. This corresponds to $\Delta \theta_\text{CCA1}=$ -4.4 (8.8) mdeg and $\Delta \theta_\text{CCA2}=$ -0.6 (-1.2) mdeg, relative to the angles for the elastic scattering. 

Once the setup in Figure~\ref{fig:setup}  is aligned, spatial and temporal overlap between the pump pulse and x-ray pulse is optimized. We assume these measurements are shot-noise limited for the purpose of the quoted errors.  For all measurements we scan only one parameter at a time, with the others nominally at their optimum value. Figure~\ref{fig:XOMdependance}~a--b shows the measured dependence of the first and second order sideband signals as a function of temporal delay between the x-ray and optical pulses. The temporal jitter between arrival of the x-ray and optical pulses had a FWHM of about 230 fs and was corrected using the arrival time monitor. Figure~\ref{fig:XOMdependance}~a is displayed on a logarithmic scale and shows multiple peaks due to the finite reflection of the optical beam inside the sample.  The secondary peaks are spaced by twice the round trip time, as the phase-matching condition is only met for the nominally co-propagating laser and x-ray.  The peak signal for the second-order sideband is about a thousand times weaker than the first-order. 
In both cases the temporal window, about 400 fs FWHM for the first order, for mixing is dominated by the non-collinear convolution of the two pulses that propagate with different group velocities $c, c/n_g$, at a crossing angle 
$\approx\theta_B$.

Figures~\ref{fig:XOMdependance}~c--d show the number of detected photons per pulse as a function of $\Delta\theta$.  The mdeg.-scale acceptance in the elastic rocking curve ( inset of Figure\ref{fig:XOMdependance} c) ) and the first order sideband shown in \ref{fig:XOMdependance} c reflect the high quality of the sample. The FWHM of the elastic Si 220 rocking curve shown in the inset of Figure \ref{fig:XOMdependance}  c) is $\sim 0.6$ mdeg. We attribute the modulation in the peak Figure \ref{fig:XOMdependance} c) to effects of the walkoff between the x-ray and optical pulse inside the crystal, which reduces the effective scattering volume. The sample detuning required to phase-match in Figures \ref{fig:XOMdependance} c--d matches the calculated values almost exactly.  
\begin{figure}[t]
\begin{center}
   \includegraphics[width=1\linewidth]{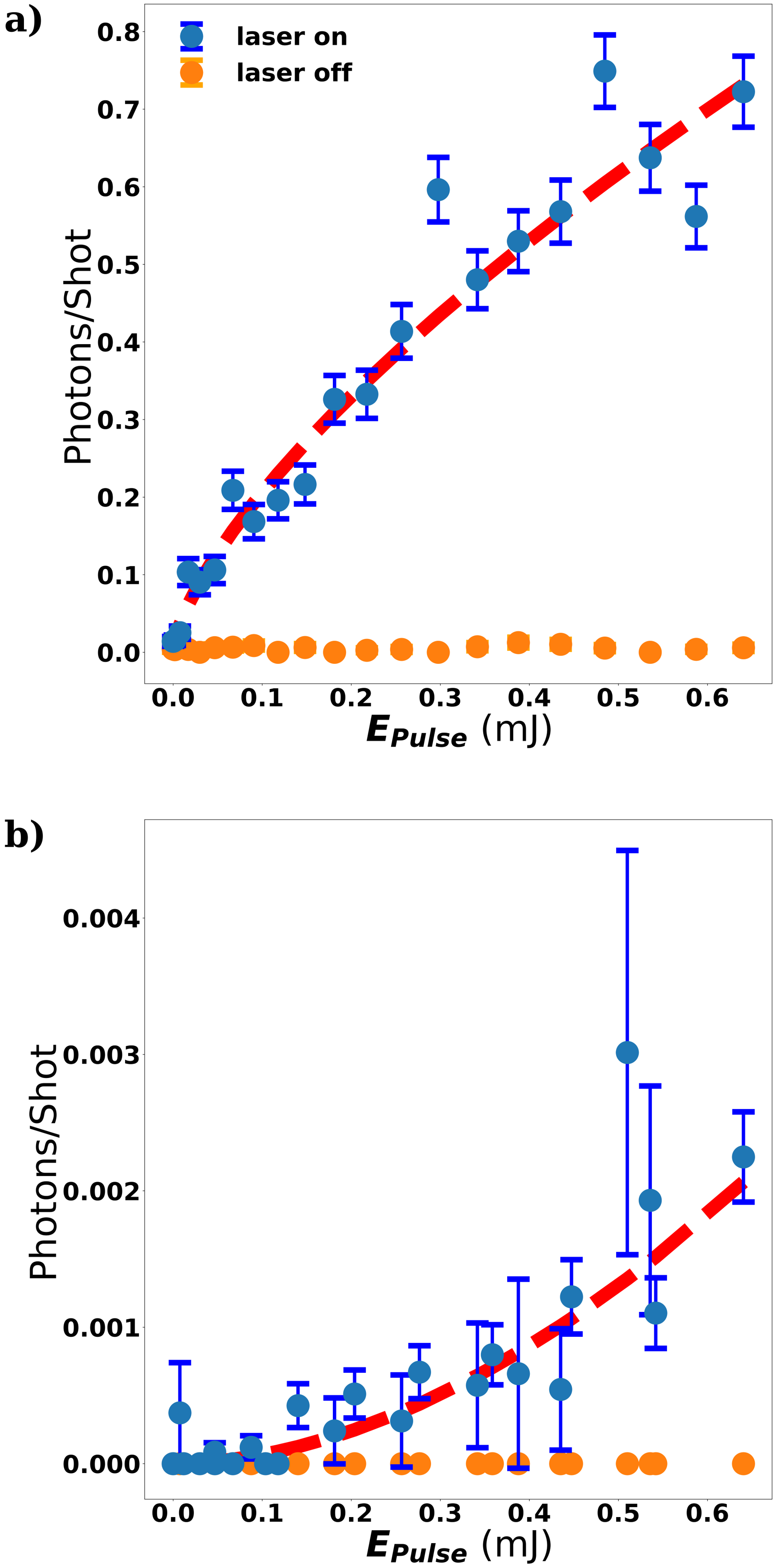}
    \caption{\label{fig:intensity}
      a) Intensity dependence of the first order sideband with fit in red.  b) The intensity dependence of the second order sideband with fit in red. 
    }
\end{center}
\end{figure}
\begin{figure*}
 \begin{center}
   \includegraphics[width=1\linewidth]{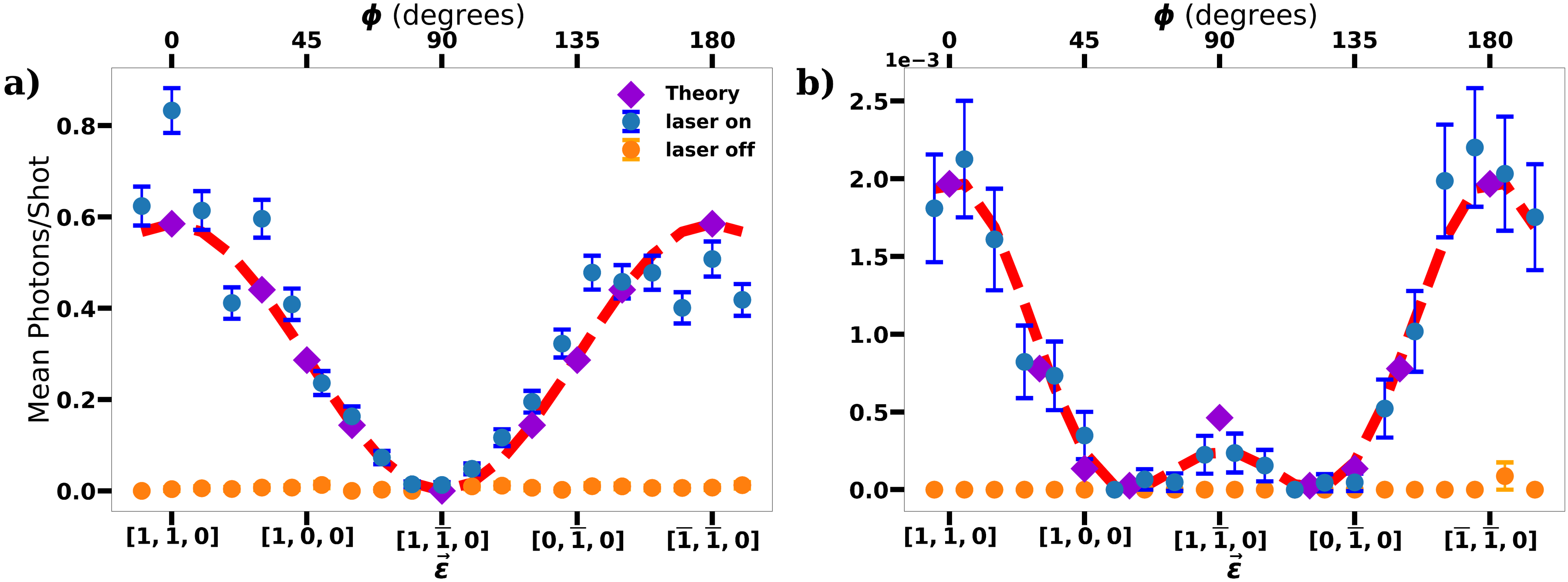}
    \caption{\label{fig:polarization}
    a) The polarization dependence of the first order sideband with fit in red. b) The polarization of second order sideband with fit in red. For both the relative theoretical values are shown in purple. }
\end{center}
\end{figure*}
Figures~\ref{fig:XOMdependance}~e--f show the number of detected photons per pulse as a function of detuned analyzer energy, $\Delta E$, at fixed $\theta_s$, about the first and second sideband respectively.  Figure~\ref{fig:XOMdependance}~e shows a well defined first-order sideband on the tails of the elastic background.  The peak position is approximately 1 eV above the elastic line (as expected),  has a signal to noise ratio of approximately 100:1 and a width of 0.2 eV (expected based on the pass band of the channel cut mono and analyzer). Similarly the second-order sideband shows an energy shift of about 2 eV, as expected for a sum frequency photon involving two pump photons, albeit with significantly lower statistics for the detected photons for both the on and off shots. Figures~\ref{fig:XOMdependance}~g--h show the detected photons per pulse as a function of $\theta_s$, at fixed $\Delta E$ corresponding to the first- and second-order sideband.  In both cases the sidebands have a width of about $\sim 1$ mdeg, comparable to the angular acceptance of the 311 Bragg peak.

The intensity of each sideband as a function of pump pulse energy are shown in Figure~\ref{fig:intensity}. In the perturbative regime, one would expect that the first-order sideband would be linearly proportional to the laser intensity (and thus pulse energy) while the second-order sideband would depend quadratically. We find that the reduced $\chi$-square for a sub-linear dependence (exponent of 0.683 $\pm$ 0.025 and $\chi$-square$/$dof of 3.228) shown in (Figure~\ref{fig:intensity}~a) corresponds to a much better fit than a linear fit (exponent of 1 and $\chi$-square$/$dof of 8.218).  The second order fits well to a nearly quadratic dependence. The deviation from the expected scaling  for the first order is significant, indicating a break-down of perturbative scaling although it remains unclear whether it is due to microscopic or macroscopic nonlinearities. 

 The measured values for the efficiency of the first- and second-order sidebands to the 220 Bragg peak are $\eta^{(1)}_{220}=3\times10^{-7}$ and $\eta^{(2)}_{220}=3 \times 10^{-10}$, relative to the (elastic) Bragg peak at the highest intensity, and when the applied field is parallel to $\vec{G}$. Here we define  $\eta^{(n)}_{\vec{G}}\equiv I^{(n)}_{\vec{G}}/I^{(0)}_{\vec{G}}$. Due to the nature of the non-collinear wave-mixing transmission geometry, $\eta^{(1)}_{220}$ and $\eta^{(2)}_{220}$ of the sidebands will be a slight underestimate because the wave-mixing only scatters from the crystal volume excited by the optical pulse, which is smaller than the crystal volume involved in the elastic scattering. We estimate a factor of 4 times less volume in the sample generating the wave-mixing photons compared to the elastically scattered photons at the Bragg peak due to walkoff between the pump and x-ray pulses. Note that the efficiency of the second-order response is much higher than what would be expected from multipolar effects, i.e. $\eta^{(2)}_{220} >>(\eta^{(1)}_{220})^2$. As discussed in Appendix \ref{a:pointdipole}, this indicates that the second-order sideband is dominated by the second-order dipole response of the valence electrons which is more sensitive to local inversion symmetry breaking. This conclusion is further supported by the measurement of the polarization dependence described below.

\begin{figure*}
    \begin{center}
        \includegraphics[width=1\linewidth]{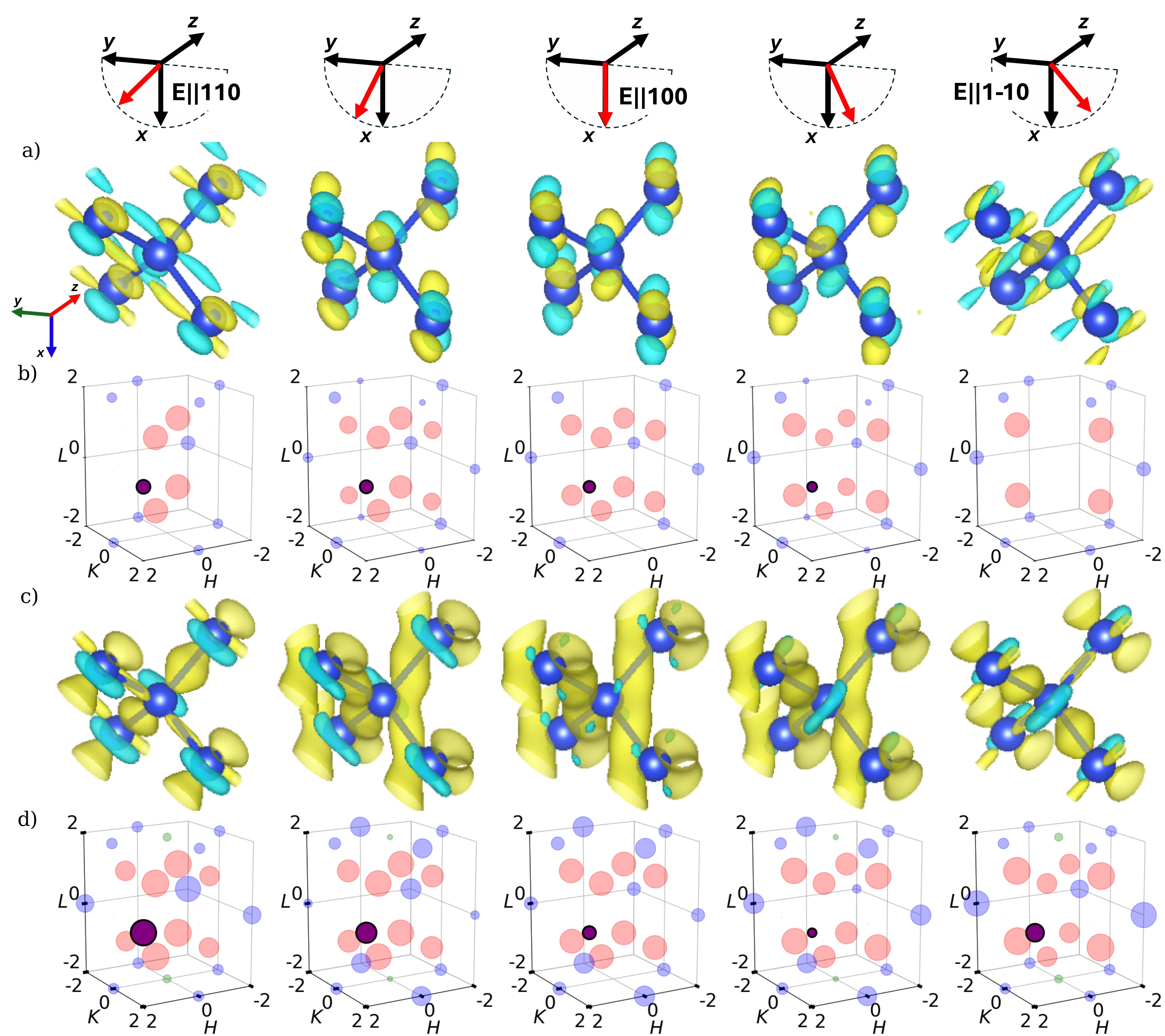}
    \\\caption{\label{fig:densities} Ab initio calculations of the induced charge densities for the first order optical response as a function of electric field direction(with the electric field shown  in red rotating within the 001 plane) in a) real space and b) reciprocal space. Corresponding ab initio calculations for induced charge densities of the second order response in c) real space and d) reciprocal space. Note that in the reciprocal space plots the red spheres correspond to points in the {111} family, the blue spheres the {220} family, and the green spheres the {200} family. The points corresponding to the 220 component in the reciprocal space plots are highlighted in purple. The size of the spheres is associated with the amplitude of that component of the charge density at the given electric field direction. The isosurfaces in a) and c)are visualized using VESTA \cite{momma2011vesta}. The yellow and blue colors represent negative and positive charges, respectively}. 
    \end{center}
\end{figure*}
 The detected photons per pulse as a function of optical electric field direction $\vec{\epsilon}$ (rotated in the plane perpendicular to (001)) is shown for the first-order sideband in Figure \ref{fig:polarization} a) along with the results of our ab initio calculations based on the Floquet-Bloch formalism\cite{popova2018theory} shown in purple (see Appendix \ref{a:calculations} for details of the calculations and Appendix \ref{a:peakchoice}.  We observe a dependence proportional to the projection of the electric field along the 220 reciprocal lattice vector, which is consistent with induced motion following the field direction. This is expected for first-order sidebands of high-symmetry reciprocal lattice vectors like the 220 and is in good agreement with our theoretical calculations.   
 Our measurement of the second-order sideband to the 220 peak as a function of the optical field direction is shown in Figure \ref{fig:polarization} b). The second-order polarization dependence is not consistent with the induced electron motion following exactly the direction of the field. This, in addition to the efficiency of the second-order sideband, is consistent with a second-order dipole response that originates from sites in the silicon unit cell with a nonzero local microscopic $\chi^{(2)}(\vec{r_s})$ or in other words where the symmetry is locally broken.  
 The form of the polarization dependence and the sensitivity of the second-order 220 component of the optical response (see Appendix \ref{a:pointdipole}) are consistent with the response originating at the sites between the atomic sites and the center of the bonds (with $C_{3v}$ site symmetry). Although the atomic sites ($T_d$ site symmetry) would be expected to also have a nonzero $\chi^{(2)}(r_s)$ individually, their contribution to the 220 component is forbidden by the symmetry of the crystal structure. The center of the bonds (the $D_{3d}$ sites) have an identically zero $\chi^{(2)}(r_s)$ because they are centers of symmetry, and do not contribute to any spatial Fourier component (For more details about the sensitivity of the 220 to different  site-symmetries see, Appendix \ref{a:peakchoice}). 

 For the $C_{3v}$ sites there are four independent second-order tensor components in terms of the cubic axes, 
($\chi_{11},\chi_{12},\chi_{14},$ and $\chi_{15}$). 
In Appendix~\ref{a:chargedensity}, we define the reduced tensor, $\Gamma^{(n)}_{\vec{G}} = \vec{G}\cdot \chi^{(n)}_{\vec{G}}$, such that $\rho^{(n)}_{\vec{G}}=\Gamma^{(n)}_{\vec{G}}\vec{E}^n$.  Thus we find that for $C_{3v}$ symmetry, $\rho^{(2)}_{220}$ the is sensitive to 3 of the 4 tensor components ( $\chi_{11},\chi_{12},$ and $\chi_{15}$) leading to a reduced tensor, 
 \begin{align}
 \Gamma^{(2)}_{220} = 8f_{32e}\sin 8\pi \epsilon\left(\begin{array}{ccc}
\chi_{11}+\chi_{12} & \chi_{15} & 0 \\
\chi_{15} & \chi_{11}+\chi_{12} & 0 \\
0 & 0 & 2\chi_{12}
 \end{array}\right) \label{e:G2-220}.
 \end{align}
For simplicity of notation we drop the 220 label for these tensor components, except for where required for clarity. In our experiment on the (001)-cut crystal, the applied electric field makes an angle $\phi$ with respect to the 220 reciprocal lattice vector and rotates in the (001) plane. For this geometry the expected polarization dependence of the second-order induced density,
\begin{equation}
          \rho^{(2)}_{220}(\vec{r}_{32e},\phi)\propto|\vec{E}|^2 (\chi^{(2)}_{11}+\chi^{(2)}_{12} +2\chi^{(2)}_{15} \cos(2\phi)). 
\end{equation}
This is consistent with what we measure in Figure \ref{fig:polarization} b). From this measurement we find that $\chi^{(2)}_{11}+\chi^{(2)}_{12} =\chi^{(2)}_{15}$. We perform additional measurements of the laser polarization-dependence of the second-order sideband to the $220$ Bragg peak in a 100 $\mu$m thick $(\bar{1}10)$ cut crystal under similar conditions. This allows us to further constrain $\chi^{(2)}_{220}$ by rotating the electric field in a plane orthogonal to the (001) face but still containing $\vec{G}=220$.  We find that $\chi^{(2)}_{11}\approx \frac{3}{2}\chi^{(2)}_{15}$, and $\chi^{(2)}_{12}\approx -\frac{1}{2}\chi^{(2)}_{15}$.  The combined results compare well with our theoretical calculations (Appendix \ref{a:calculations} and are given in table \ref{t:results}. 

While we have only measured two sidebands about a single Bragg peak, the very good qualitative and quantitative agreement gives us confidence that our formalism will be able to predict the full excited-state density similarly well. 
In figure \ref{fig:densities}, we show the predicted first and second order induced density in both real (a and c) and reciprocal space (b and d) (for the $\{200\},\{111\},
\{220\}$ family of planes) for a representative set of applied electric field directions in the (001) plane and at fixed field strength of $5\times 10^{11}$ W/cm$^2$.  For the reciprocal space plots, the amplitude is encoded in the size of the spheres, and the color is used to encode the different families of planes. The real space density is limited to the valence density, and appears primarily off the atomic sites (This is to be expected since our first-principles calculations are based on the pseudo-potential method and only include the top four valence bands  originating from 3$s$ and 3$p$ electrons).  The first-order density largely follows the field as seen in our measurements. Note that the applied field breaks the cubic symmetry of the crystal (dynamically), such that different Fourier components from the same family are in general no longer equivalent. Moreover, while the first-order response in real space is rather complicated, $\rho^{(1)}_{\{111\}}$ and $\rho^{(1)}_{\{220\}}$ are proportional to $(\vec{G}\cdot\vec{E})^2$.  The second-order response is even more nontrivial and clearly not proportional to $(\vec{G}\cdot\vec{E})^4$, as in the experimental case for the $220$. In addition, we see that $002$, is allowed for $\vec{E}$ having a component along both $\hat{x}$ and $\hat{y}$ directions. This suggests that the field breaks the glide plane symmetry, which ordinarily is responsible for the strong-forbidden nature of the $\{00l\}$ family of planes when $l=4m-2$. 

\begin{table}

\begin{tabular}{| p{2.3cm}|p{2.3cm}|p{2.3cm} |}
  \hline
    & Measurement & Theory\\
 \hline
$\chi^{(2)}_{11}+\chi^{(2)}_{12}$ &   $1.0\pm 0.2$& $0.7$\\
 $\chi^{(2)}_{11}$                 & $1.5\pm 0.2$& $1.5$\\
 $\chi^{(2)}_{12} $                &$-0.6\pm 0.1$& -0.83\\
 \hline
\end{tabular}
\caption{Extracted Components of $\chi^{(2)}_{220}$ in units of 
       $\chi^{(2)}_{15}$
\label{t:results}}
\end{table}

\section{Conclusions and Future Directions\label{s:conclusions}}

We have reported measurements of x-ray and optical wavemixing in silicon excited by below bandgap optical photons. We have used phase-matching and crystal optics to extract the weak nonlinear sum-frequency signals corresponding to mixing of single x-ray photons with one and two optical photons in single-crystal silicon. In principle atomic-scale movies of the electron motion within the unit cell could be constructed by measuring multiple sidebands to multiple Bragg peaks and determining their relative phase (for example using maximum entropy methods or multibeam diffraction). Nonetheless, we have shown here that the polarization dependence of a single second-order sideband contains sufficient information to state that the second-order response is dominated by dipole effects allowed by local inversion symmetry breaking with negligible contribution from higher-order multipoles. 
Furthermore, we are able to attribute the second-order dipole response to particular non-centrosymmetric high-symmetry ($C_{3v}$) sites along the $sp^3$ bonds within a phenomenological point-dipole model (see Appendix C) and without direct phase information. We were also able to determine three of the four symmetry equivalent components of (the Fourier transform) of the local second-order optical susceptibility tensor, constrained by the particular Bragg peak. The results agree well with first-principles, Bloch-Floquet calculations for the optically excited valence electron density. 
These findings have important implications for understanding electronic structure and dynamics in both near equilibrium and strongly-driven materials. For example XOM may help disentangle the relative contributions of strongly-driven intra- and inter-band currents and the role of recollisions on solid-state high-harmonic generation\cite{YouMgO,GhimireAndReisReview} or how light-induced changes in electronic structure lead to novel functionality in quantum materials \cite{Basov}.

Here new methods are required to move beyond perfect crystals (allowing this measurement in a much wider range of materials), and ideally to perform the measurements using time-domain scattering with newly available attosecond hard x-ray pulses\cite{LCLS1stHardAtto,EUXFELAtto,attovision}.

\section{Acknowledgements}
This work is supported by the US Department of Energy, Office of Science, Office of Basic Energy Sciences, Chemical Sciences, Geosciences, and Biosciences Division through the AMOS program and performed on the LCLS. We thank X. Huang for the fabrication and advice on the design of the Si 311 channel cut crystals. We thank Priyanka Chakraborti, Ozgur Culfa, Johann Haber, Samuel Teitelbaum, Taito Osaka, Tadashi Togashi, Yuichi Inubushi , Ichiro Inoue , Keiichi Sueda, Jumpei Yamada for help with preparatory experiments on SACLA (2019B8062, 2018B8086) and Christoph Sahle for letting us borrow Si 400 channel cut crystals. We thank S. Gerber, Nelson Hua, Ludmila Diniz Leroy, H. Lemke, R. Mankowsky, M. Sander and C. Svetina for help with preliminary experiments on SwissFEL.  Use of the Linac Coherent Light Source (LCLS), SLAC National Accelerator Laboratory, is supported by the U.S. Department of Energy, Office of Science, Office of Basic Energy Sciences under Contract No. DE-AC02-76SF00515. T.B. acknowledges the funding by the Cluster of Excellence ``CUI: Advanced Imaging of Matter'' of the Deutsche Forschungsgemeinschaft (DFG) -- EXC 2056 -- project ID 390715994. D. P.-G. acknowledges the funding from the Volkswagen Foundation through a Freigeist Fellowship, grant number 96 237. S. Shwartz acknowledges the support of the Israel Science Foundation (grant No. 847/21). We acknowledge valuable comments by Robin Santra.

\appendix

\section{Optically induced-charge density\label{a:chargedensity}}

We take the total microscopic electron density, $\rho(\vec{r},t)= \rho_0(\vec{r})+\Delta \rho(\vec{r}, t)$ where $\rho_0(\vec{r})$ is the field-free density (valence and core) that is periodic in space $\rho(\vec{r}+\vec{R})=\rho(\vec{r})$ for all lattice vectors $\vec{R}$  and $\Delta \rho (\vec{r}, t)$ is the time-dependent charge (valence) density induced by the low-frequency optical (IR) electric field $\vec{E}(\vec{r}, t)$ in the presence of screening and local-field effects. 

We approximate the optical excitation as a monochromatic plane wave with period $T_o$ and angular frequency $\omega_o= 2\pi/T_o$.  The charge density  induced by the excitation field corresponds to an incommensurate and traveling charge density wave, and thus $\Delta\rho(\vec{r}, t)=\Delta\rho(\vec{r}+\vec{R}, t+T_o)$ up to a relatively slowly varying phase factor. We expand the total charge density in terms of its Fourier components,
\begin{equation}
\rho(\vec{r},t)=\sum_{\vec{G}, n}\rho(\vec{G}+n \vec{k}_o, n \omega_o) e^{-i((\vec{G}+n \vec{k_o})\cdot \vec{r} - n \omega_o t)} 
\end{equation}
where $\vec{G}$ are the reciprocal lattice vectors, $n$ are harmonics of the optical frequency, $k_o= 2 \pi n_o/\lambda_o$  is the wavevector of the optical pump with vacuum wavelength $\lambda_o$ and index of refraction $n_o$. For simplicity of notation, we define $\rho^{(n)}_{\vec{G}} \equiv \rho(\vec{G}+n\vec{k}_o,n\omega_o)\approx\rho(\vec{G},n\omega_o)$ since $\lambda_o$ is long compared to the lattice spacing.

\section{Ab-initio calculations for silicon\label{a:calculations}}
We perform ab initio calculations of the induced charge density, $\rho^{(n)}(\vec{r})$, for optically excited silicon within the Floquet-Bloch formalism \cite{popova2024microscopic}. 
We consider a driving field with a photon energy of 0.95 eV and an intensity of $5\times 10^{11}$ W/cm$^2$. 
The calculations are performed using 4 valence bands and 76 conduction bands on a four-times shifted $12\times 12\times 12$ Monkhorst-Pack $\mathbf{k}$-point grid. The infinite Floquet-Bloch Hamiltonian is approximated by a matrix with 301 blocks, each containing 80 states. The $\mathbf{k}$-point grid, the number of bands and the blocks are selected based on a convergence study. We apply the scissors approximation \cite{levine1989linear} to correct the direct band gap from the calculated 2.5 eV to the experimental value of 3.5 eV \cite{phillips1962band, hybertsen1985first}. 

We define the reduced tensor $\Gamma^{(n)}(\vec{r})=\nabla\cdot\chi^{(n)}(\vec{r})$, in terms of a local nonlinear susceptibility, $\chi^{(n)}(\vec{r})$, such that, 
\begin{equation}
    \rho^{(n)}(\vec{r})=\Gamma^{(n)}(\vec{r}) \cdot [\vec{E}]^n. 
\end{equation}
Thus, from the reduced tensor, we can predict the induced charge density, oscillating at the harmonics of the optical excitation for a given field magnitude and direction, within the limits of perturbation theory.  In the case that $n=0$, we recover the ground state valence density, (excluding the core electron density, since we only consider the highest four valence bands). 

\begin{figure}
\begin{center}

   \includegraphics[width=1.0\linewidth]{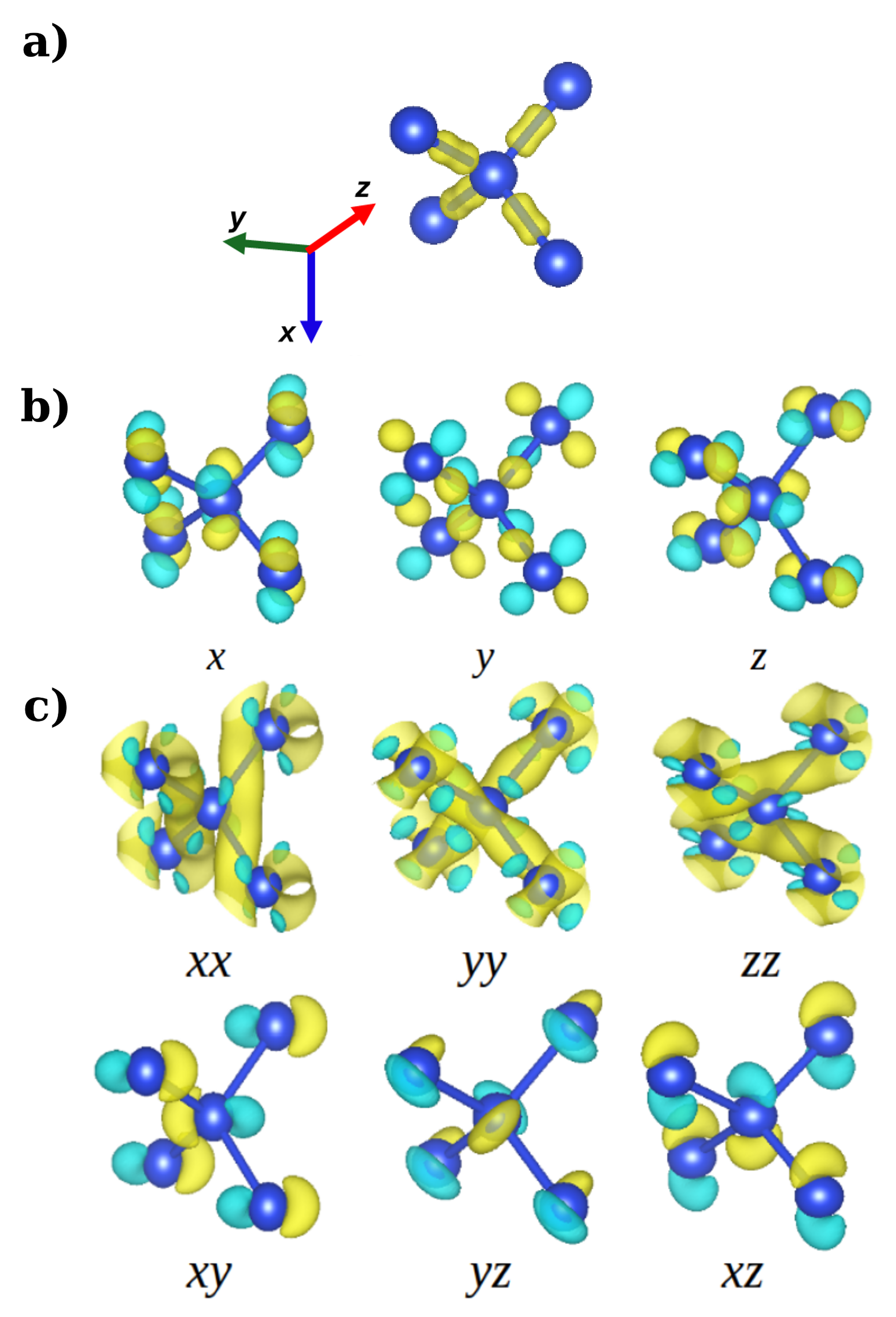}
    \caption{\label{fig:Tensor}Ab initio calculations of the reduced tensor  components of the valence density in real space for a) the 0th, b) the first, and c) the second-order response to the optical field.     Calculations performed for an optical intensity of $5\times 10^{11}$ W/cm$^2$. }
\end{center}
\end{figure}

In Figure~\ref{fig:Tensor}, we show the results of our calculations of $\Gamma^{(n)}(\vec{r})$ for the components of the a) ground state valence density ($n=0$), b) its first-order response ($n= 1$), and c) its second-order response ($n= 2$).  
Note that the ground-state valence density, $\Gamma^{(0)}(\vec{r})$, is distributed about the center of the bonds, as expected. The first-order reduced tensor components, $\Gamma_i^{(1)}(\vec{r})$, where $i=x,y,z$ along the cubic directions,  show the density distributed about the tetrahedral sites while the second order reduced tensor components, $\Gamma_{ij}^{(2)}(\vec{r})$, show the second order density shift off the center of the bond towards the atomic sites but not onto the atomic sites. 
The measurements of the second order response will turn out to be consistent with optically induced valence density in the space between the atomic sites and the center of the bonds. 

In the experiments, we measure individual spatial and temporal Fourier components of the induced density.  These can be predicted from the spatial Fourier components of $\Gamma^{(n)}(\vec{r})$,
\begin{equation}
\Gamma^{(n)}_{\vec{G}} =\vec{G}\cdot \chi^{(n)}_{\vec{G}},
\end{equation}
where $\chi^{(n)}_{\vec{G}}$ are the spatial Fourier components of $\chi^{(n)}(\vec{r})$. 
Thus, the Fourier components of the charge density,
\begin{equation}
    \rho^{(n)}_{\vec{G}}= \Gamma^{(n)}_{\vec{G}} \cdot [\vec{E}]^{n} .
\end{equation}
We note that with this definition the ground-state charge density is defined as $\rho^{(0)}(\vec{r})=\Gamma^{(0)}(\vec{r})=\nabla\cdot E_{loc.}(\vec{r})$ where $E_{loc.}(\vec{r})$ is the  local field inside the crystal (in absence of any external perturbation).  In Section \ref{s:results}, we present results for the first- and second-order induced density in both real and reciprocal space (for the $\{200\},\{111\},\\
\{220\}$ family of planes) for a representative set of applied electric field directions sampled in the experiment, and at fixed magnitude.

In the kinematic limit\cite{warren1990x}, the intensity of a sideband to a reciprocal lattice vector $\vec{G}$,
\begin{equation}
    I^{(n)}_{\vec{G}} \propto|\rho^{(n)}_{\vec{G}}|^2=|\Gamma^{(2)}_{\vec{G}}\cdot [\vec{E}]^n|^2 . 
\end{equation}
We define the efficiency of a given sideband relative to the elastic Bragg peak, 
\begin{equation}
    \eta^{(n)}_{\vec{G}}= \frac{I^{(n)}_{\vec{G}}}{I^{(0)}_{\vec{G}}}.
\end{equation}
Thus up to macroscopic propagation effects including walkoff, $\eta^{(n)}_{\vec{G}} = (\rho^{(n)}_{\vec{G}}/\rho^{(0)}_{\vec{G}})^2$.

\section{Point Dipole Model\label{a:pointdipole}}
To complement the  ab initio calculations we present a phenomenological point-dipole model of the optically induced charge density. We use this minimal model both to provide an intuitive interpretation of the measurements based on symmetry and facilitate the connection between our measurements and our ab initio calculations. In this model we approximate the time- varying charge density, analogous to the independent atom approximation, as
    \begin{equation}
        \rho(\vec{r},t)=\sum_s f_s\delta(\vec{r}-\vec{r}_s -\vec{u}_s(t)),
    \end{equation}
 where $f_s$ is the electron density at site $s$, and  $\vec{u}_s(t)=\sum_n u_s^{(n)}\text{exp}(in\omega_ot)$ is the induced displacement about its equilibrium position, $\vec{r}_s$, oscillating at harmonics of the drive frequency. Thus, 
\begin{align}
    \rho^{(n)}_{\vec{G}} = \frac{1}{T_oV}\int\text{d}t \text{e}^{i\omega t}\sum_s
    f_s\text{e}^{i\vec{G}\cdot\vec{r}_s}\text{e}^{i\vec{G}\cdot\vec{u}_s(t)}
\end{align}
 Note that even if $\vec{u}(t)$  oscillates only at the the fundamental, $\omega_0$, the Fourier components of the induced density will still contain both the even and odd harmonics since the displacements appear in the exponential.  For example,  in this case, $\rho^{(2)}\approx \frac{1}{2}(\rho^{(1)})^2$.  
 In the experiments, we find that $\rho^{(2)}_{220}\gg (\rho^{(1)})^2$, and has a very different polarization dependence.
 Thus, $u^{(2)}_{220}\neq 0$, and the leading-order the nonlinear response can be approximated as a local nonlinear dipole,   
$\vec{p}_s^{(n)}=-f_s \vec{u}_s^{(n)}$.  In analogy with macroscopic nonlinear optics, we expand the dipole moments for site s as 
 \begin{equation}
     \vec{p}_s^{(n)}= \chi^{(n)}(\vec{r}_s)[\vec{E}]^{n}. 
 \end{equation}
We note, unlike in macroscopic nonlinear optics, the local susceptibility $\chi^{(n)}(\vec{r}_s)$ depends on the site symmetry about position $r_s$ within the unit cell of the crystal, and thus even order nonlinearities can arise due to local inversion symmetry breaking in centrosymmetric materials, for $\vec{G}\neq0$.

\section{Wave-mixing sensitivity in the silicon structure\label{a:peakchoice}}
 \begin{figure}[b]
    \centering
    \includegraphics[width=1\linewidth]{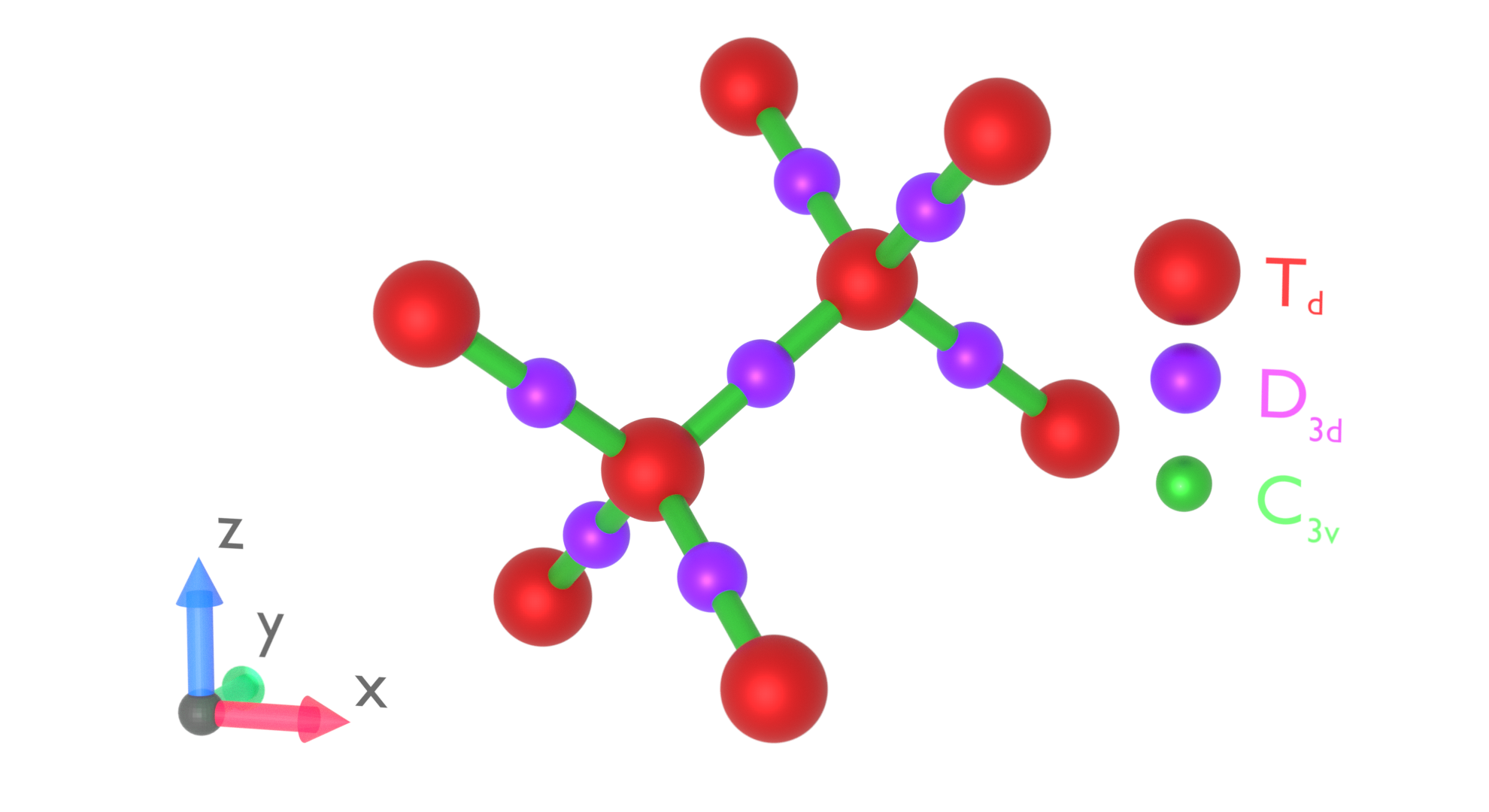}
    \caption{An illustration of the high symmetry sites along the the tetrahedral bonds for silicon. The Red spheres represent the atomic positions at the 8a Wykoff Positions with $T_d$ symmetry, the purple spheres represent the $D_{3d}$ sites at the center of each bond, and the green between the $T_d$ and $D_{3d}$ sites represents the $C_{3v}$ sites.}
   \label{fig:sites}
\end{figure}

 We note that silicon is isostructural with diamond, sharing the same high-symmetry $m\bar{3}m (O_h)$ centrosymmetric point group and Fd$\bar{3}$m space group. There are 8 atoms per conventional face-center-cubic unit cell occupying the (8a) tetrahedral sites with $\bar{4}3m\ (T_d)$ symmetry.
A portion of the structure, showing the local coordination is shown in Figure \ref{fig:sites}.  Here the tetrahedral atomic sites are shown in red, and the covalent bonds are shown schematically in green.  Half way between each bonds are centers of symmetry corresponding to the 16c sites with $\bar{3}m\  (D_{3d})$ site symmetry;  while a general position along the bonds corresponds to the 32e sites with $3m\ (C_{3v})$ site symmetry.  While the overall structure has inversion symmetry, clearly both the atomic sites and the general positions along the bond lack  inversion symmetry about their respective sites.
 
Note that diamond structure $|\rho_{hkl}|\neq 0$ for $h,k,l$ odd, or $h+k+l=4m$, for $h,k,l$ all even and $m$ and integer.  For point charges, $|\rho_{hkl}|$ is maximal ($8f/a^3)$ for the $h+k+l=4m$ case such as the 220, as the tetrahedral sites lie within the corresponding atomic planes (for suitable choice of origin). As the valence charge distribution is non-spherically distributed about these sites, it gives rise to finite $\rho_{hkl}$ corresponding to certain otherwise forbidden Bragg peaks such as the 222, allowing for weak diffraction from this charge density. 

Due to the cubic symmetry of the lattice, the macroscopic linear optical properties are isotropic, and the induced macroscopic polarization follows the applied field.
While this is not strictly true for the microscopic linear response, $\rho^{(1)}_{\vec{G}}\propto\vec{G}\cdot \vec{E}$
for certain $\vec{G}$ along high symmetry directions (e.g the 111, 220).   In contrast, the local nonlinear response can be anisotropic, even along these high symmetry directions. 

The form of $\chi^{(n)}_{\vec{(G)}}$ and thus $\Gamma^{(n)}_{\vec{G}}$ depend on the local symmetries of $\chi^{(n)}(\vec{r})$ and the particular $\vec{G}$.
It can be readily seen that $\Gamma^{(2)}_{hkl}=0$ for the tetrahedral sites, when $h+k+l=4m$, even though locally these sites break inversion symmetry.
This is due to the overall inversion symmetry of the point group that requires $\chi^{(2)}(\vec{r})=-\chi^{(2)}(-\vec{r})$ such that the second-order dipoles at the tetrahedral sites are 180 degrees out of phase in every other plane.   Thus, as we show here,  by an analysis of the second-order sidebands from the $220$ reflection, we can localize the induced charge away from the tetrahedral sites by its site symmetry.

\section{Isolation of Fourier Components\label{a:phasematching}}

To isolate a particular spatial and temporal Fourier component of the induced density, we make use of phase-matching in combination with an energy-resolving analyzer. Although the optical wavevector is ten-thousand times smaller than the x-ray wavevector, it plays an important role in the phase-matching condition.
 In general, phase matching  is accomplished by satisfying momentum and energy conservation,
\begin{align}
 \vec{k}^{(n)}-\vec{k}_x = n\vec{k}_o + \vec{G} \equiv \vec{Q}^{(n)}, \text{ and} \\
 \omega^{(n)}=\omega_x+n\omega_o, 
\end{align}
where $\vec{k}_o,\vec{k}_x,\vec{k}^{(n)}$ are the optical, incident x-ray, and nonlinear diffracted x-ray wavevectors inside the material. 
The angle between the incident and diffracted x rays, $\theta_s$ satisfies 
\begin{equation}
\cos\theta_s \equiv \hat{k}_x\cdot\hat{k}^{(n)} = \frac{({k}^{(n)})^{ 2}+k_x^2-(Q^{(n)})^2}{2k_x k^{(n)}}.
\end{equation}
While the angle between the incident x rays and the lattice planes,
\begin{equation}\begin{split}
\theta = \sin^{-1}\left( \frac{Q^{(n)}\sin\theta_{Q^{(n)}}}{\sqrt{ (G+k_{o\parallel})^2 + k_o^2-k^2_{o\perp}}} \right) + \\
\frac{n}{|n|} \tan^{-1}\left(\frac{k_{o\parallel}}{G+k_{o\parallel}}\right), 
\end{split}
\end{equation}
where $k_{o\parallel}$, and $k_{o\perp}$  are the components of  $\vec{k}_o$  within and perpendicular to the plane containing G and $k_x$, respectively.  
\begin{equation}
\sin\theta_Q \equiv -\hat{k}_x\cdot\hat{Q}^{(n)} = \frac{(Q^{(n)})^{2}-(k^{(n)})^{2}+k_x^2}{2 k_x Q^{(n)} },
\end{equation}
is the sine of the angle the incident x rays make relative to the planes perpendicular to $\vec{Q}^{(n)}$.

Since  $\omega_o \ll \omega_x, k_o \ll k_x$, to first order in the ratio $x =\omega_o/\omega_x$, the deviation of the input angle relative to the Bragg angle $\theta_B$ is
\begin{align}
\Delta\theta & \equiv \theta-\theta_B \\
&  = \frac{n x}{\sin2\theta_B}\left(n_o(\cos\beta\cos\theta_G-\sin\alpha\sin\theta_B )-1\right). 
\end{align}
where, $n_o$ is the index of refraction of the optical beam. $\alpha$ is the angle that the internal optical beam makes relative to the lattice planes and  $\gamma$ is the angle that the optical beam makes perpendicular to the plane containing $\vec{k}$ and $\vec{G}$ (note that $\cos^2\beta = \cos^2\alpha-\sin^2\gamma$). We have neglected the small difference in the x-ray index from unity.  Thus if $\vec{q}$ is also largely in plane, such that $\alpha \approx \beta$, 
\begin{align}
\Delta\theta = \frac{n x}{\sin2\theta_B}\left(n_o\cos(\theta_B+\alpha)-1\right). 
\end{align}
Note that the deviation of the scattering angle from $2\theta_B$, 
\begin{align}
\Delta\theta_s & \equiv  \theta_s - 2\theta_B \\
&\approx -\frac{n x}{\cos\theta_B} (n_o\sin\alpha + \sin\theta_B),
\end{align}
independent of $\gamma$.  Since $x \ll 1$, $\Delta\theta, \Delta\theta_s \ll 1$, the difference between the exact solution and the first-order approximation is negligible. Nonetheless, since the induced density is a small fraction of the total density, we rely on the finite deviations to isolate the mixing signal.  This imposes stringent requirements on the monochromitization and collimation of the incident beam as well as the background rejection that includes rejecting the tails of the elastic Bragg peak. This is true even for measurements of wave-mixing in nominally perfect crystals such as diamond and silicon.  As described in section \ref{s:setup}, we achieve this by using a matched pair of Si (311) channel-cut crystals in a dispersive geometry for both the monochromator and analyzer.

\bibliography{references_v1}

\begin{thebibliography}{33}%
\makeatletter
\providecommand \@ifxundefined [1]{%
 \@ifx{#1\undefined}
}%
\providecommand \@ifnum [1]{%
 \ifnum #1\expandafter \@firstoftwo
 \else \expandafter \@secondoftwo
 \fi
}%
\providecommand \@ifx [1]{%
 \ifx #1\expandafter \@firstoftwo
 \else \expandafter \@secondoftwo
 \fi
}%
\providecommand \natexlab [1]{#1}%
\providecommand \enquote  [1]{``#1''}%
\providecommand \bibnamefont  [1]{#1}%
\providecommand \bibfnamefont [1]{#1}%
\providecommand \citenamefont [1]{#1}%
\providecommand \href@noop [0]{\@secondoftwo}%
\providecommand \href [0]{\begingroup \@sanitize@url \@href}%
\providecommand \@href[1]{\@@startlink{#1}\@@href}%
\providecommand \@@href[1]{\endgroup#1\@@endlink}%
\providecommand \@sanitize@url [0]{\catcode `\\12\catcode `\$12\catcode `\&12\catcode `\#12\catcode `\^12\catcode `\_12\catcode `\%12\relax}%
\providecommand \@@startlink[1]{}%
\providecommand \@@endlink[0]{}%
\providecommand \url  [0]{\begingroup\@sanitize@url \@url }%
\providecommand \@url [1]{\endgroup\@href {#1}{\urlprefix }}%
\providecommand \urlprefix  [0]{URL }%
\providecommand \Eprint [0]{\href }%
\providecommand \doibase [0]{https://doi.org/}%
\providecommand \selectlanguage [0]{\@gobble}%
\providecommand \bibinfo  [0]{\@secondoftwo}%
\providecommand \bibfield  [0]{\@secondoftwo}%
\providecommand \translation [1]{[#1]}%
\providecommand \BibitemOpen [0]{}%
\providecommand \bibitemStop [0]{}%
\providecommand \bibitemNoStop [0]{.\EOS\space}%
\providecommand \EOS [0]{\spacefactor3000\relax}%
\providecommand \BibitemShut  [1]{\csname bibitem#1\endcsname}%
\let\auto@bib@innerbib\@empty
\bibitem [{\citenamefont {Ashcroft}\ and\ \citenamefont {Mermin}(1976)}]{AschcroftMermin}%
  \BibitemOpen
  \bibfield  {author} {\bibinfo {author} {\bibfnamefont {N.~W.}\ \bibnamefont {Ashcroft}}\ and\ \bibinfo {author} {\bibfnamefont {N.~D.}\ \bibnamefont {Mermin}},\ }\href@noop {} {\emph {\bibinfo {title} {{S}olid {S}tate {P}hysics}}}\ (\bibinfo  {publisher} {Holt-Saunders},\ \bibinfo {year} {1976})\BibitemShut {NoStop}%
\bibitem [{\citenamefont {Martin}(1968)}]{Martin1968}%
  \BibitemOpen
  \bibfield  {author} {\bibinfo {author} {\bibfnamefont {R.~M.}\ \bibnamefont {Martin}},\ }\bibfield  {title} {\bibinfo {title} {Lattice vibrations in silicon: Microscopic dielectric model},\ }\href {https://doi.org/10.1103/PhysRevLett.21.536} {\bibfield  {journal} {\bibinfo  {journal} {Phys. Rev. Lett.}\ }\textbf {\bibinfo {volume} {21}},\ \bibinfo {pages} {536} (\bibinfo {year} {1968})}\BibitemShut {NoStop}%
\bibitem [{\citenamefont {Martin}(1969)}]{Martin1969}%
  \BibitemOpen
  \bibfield  {author} {\bibinfo {author} {\bibfnamefont {R.~M.}\ \bibnamefont {Martin}},\ }\bibfield  {title} {\bibinfo {title} {Dielectric screening model for lattice vibrations of diamond-structure crystals},\ }\href {https://doi.org/10.1103/PhysRev.186.871} {\bibfield  {journal} {\bibinfo  {journal} {Phys. Rev.}\ }\textbf {\bibinfo {volume} {186}},\ \bibinfo {pages} {871} (\bibinfo {year} {1969})}\BibitemShut {NoStop}%
\bibitem [{\citenamefont {James}(1948)}]{James}%
  \BibitemOpen
  \bibfield  {author} {\bibinfo {author} {\bibfnamefont {R.~W.}\ \bibnamefont {James}},\ }\bibfield  {title} {\bibinfo {title} {The optical principles of the diffraction of x-rays},\ }\href@noop {} {\  (\bibinfo {year} {1948})}\BibitemShut {NoStop}%
\bibitem [{\citenamefont {Roberto}\ \emph {et~al.}(1974)\citenamefont {Roberto}, \citenamefont {Batterman},\ and\ \citenamefont {Keating}}]{Batterman222}%
  \BibitemOpen
  \bibfield  {author} {\bibinfo {author} {\bibfnamefont {J.~B.}\ \bibnamefont {Roberto}}, \bibinfo {author} {\bibfnamefont {B.~W.}\ \bibnamefont {Batterman}},\ and\ \bibinfo {author} {\bibfnamefont {D.~T.}\ \bibnamefont {Keating}},\ }\bibfield  {title} {\bibinfo {title} {Diffraction studies of the (222) reflection in ge and si: Anharmonicity and the bonding electron},\ }\href {https://doi.org/10.1103/PhysRevB.9.2590} {\bibfield  {journal} {\bibinfo  {journal} {Phys. Rev. B}\ }\textbf {\bibinfo {volume} {9}},\ \bibinfo {pages} {2590} (\bibinfo {year} {1974})}\BibitemShut {NoStop}%
\bibitem [{\citenamefont {Hastings}\ and\ \citenamefont {Batterman}(1975)}]{hastings1975high}%
  \BibitemOpen
  \bibfield  {author} {\bibinfo {author} {\bibfnamefont {J.}~\bibnamefont {Hastings}}\ and\ \bibinfo {author} {\bibfnamefont {B.}~\bibnamefont {Batterman}},\ }\bibfield  {title} {\bibinfo {title} {High-order anharmonic forbidden neutron reflections in silicon},\ }\href@noop {} {\bibfield  {journal} {\bibinfo  {journal} {Physical Review B}\ }\textbf {\bibinfo {volume} {12}},\ \bibinfo {pages} {5580} (\bibinfo {year} {1975})}\BibitemShut {NoStop}%
\bibitem [{\citenamefont {Yang}\ and\ \citenamefont {Coppens}(1974)}]{Yangandcoppens}%
  \BibitemOpen
  \bibfield  {author} {\bibinfo {author} {\bibfnamefont {Y.}~\bibnamefont {Yang}}\ and\ \bibinfo {author} {\bibfnamefont {P.}~\bibnamefont {Coppens}},\ }\bibfield  {title} {\bibinfo {title} {On the experimental electron distribution in silicon},\ }\href@noop {} {\bibfield  {journal} {\bibinfo  {journal} {Solid State Communications}\ }\textbf {\bibinfo {volume} {15}},\ \bibinfo {pages} {1555} (\bibinfo {year} {1974})}\BibitemShut {NoStop}%
\bibitem [{\citenamefont {Lu}\ \emph {et~al.}(1993)\citenamefont {Lu}, \citenamefont {Zunger},\ and\ \citenamefont {Deutsch}}]{LuZungerDeuthsch}%
  \BibitemOpen
  \bibfield  {author} {\bibinfo {author} {\bibfnamefont {Z.~W.}\ \bibnamefont {Lu}}, \bibinfo {author} {\bibfnamefont {A.}~\bibnamefont {Zunger}},\ and\ \bibinfo {author} {\bibfnamefont {M.}~\bibnamefont {Deutsch}},\ }\bibfield  {title} {\bibinfo {title} {Electronic charge distribution in crystalline diamond, silicon, and germanium},\ }\href {https://doi.org/10.1103/PhysRevB.47.9385} {\bibfield  {journal} {\bibinfo  {journal} {Phys. Rev. B}\ }\textbf {\bibinfo {volume} {47}},\ \bibinfo {pages} {9385} (\bibinfo {year} {1993})}\BibitemShut {NoStop}%
\bibitem [{\citenamefont {Freund}\ and\ \citenamefont {Levine}(1970)}]{Freund}%
  \BibitemOpen
  \bibfield  {author} {\bibinfo {author} {\bibfnamefont {I.}~\bibnamefont {Freund}}\ and\ \bibinfo {author} {\bibfnamefont {B.}~\bibnamefont {Levine}},\ }\bibfield  {title} {\bibinfo {title} {Optically modulated x-ray diffraction},\ }\href@noop {} {\bibfield  {journal} {\bibinfo  {journal} {Physical Review Letters}\ }\textbf {\bibinfo {volume} {25}},\ \bibinfo {pages} {1241} (\bibinfo {year} {1970})}\BibitemShut {NoStop}%
\bibitem [{\citenamefont {Eisenberger}\ and\ \citenamefont {McCall}(1971)}]{Eisenberger}%
  \BibitemOpen
  \bibfield  {author} {\bibinfo {author} {\bibfnamefont {P.}~\bibnamefont {Eisenberger}}\ and\ \bibinfo {author} {\bibfnamefont {S.}~\bibnamefont {McCall}},\ }\bibfield  {title} {\bibinfo {title} {Mixing of x-ray and optical photons},\ }\href@noop {} {\bibfield  {journal} {\bibinfo  {journal} {Physical Review A}\ }\textbf {\bibinfo {volume} {3}},\ \bibinfo {pages} {1145} (\bibinfo {year} {1971})}\BibitemShut {NoStop}%
\bibitem [{\citenamefont {Glover}\ \emph {et~al.}(2012)\citenamefont {Glover}, \citenamefont {Fritz}, \citenamefont {Cammarata}, \citenamefont {Allison}, \citenamefont {Coh}, \citenamefont {Feldkamp}, \citenamefont {Lemke}, \citenamefont {Zhu}, \citenamefont {Feng}, \citenamefont {Coffee} \emph {et~al.}}]{Glover}%
  \BibitemOpen
  \bibfield  {author} {\bibinfo {author} {\bibfnamefont {T.~E.}\ \bibnamefont {Glover}}, \bibinfo {author} {\bibfnamefont {D.}~\bibnamefont {Fritz}}, \bibinfo {author} {\bibfnamefont {M.}~\bibnamefont {Cammarata}}, \bibinfo {author} {\bibfnamefont {T.}~\bibnamefont {Allison}}, \bibinfo {author} {\bibfnamefont {S.}~\bibnamefont {Coh}}, \bibinfo {author} {\bibfnamefont {J.}~\bibnamefont {Feldkamp}}, \bibinfo {author} {\bibfnamefont {H.}~\bibnamefont {Lemke}}, \bibinfo {author} {\bibfnamefont {D.}~\bibnamefont {Zhu}}, \bibinfo {author} {\bibfnamefont {Y.}~\bibnamefont {Feng}}, \bibinfo {author} {\bibfnamefont {R.}~\bibnamefont {Coffee}}, \emph {et~al.},\ }\bibfield  {title} {\bibinfo {title} {X-ray and optical wave mixing},\ }\href@noop {} {\bibfield  {journal} {\bibinfo  {journal} {Nature}\ }\textbf {\bibinfo {volume} {488}},\ \bibinfo {pages} {603} (\bibinfo {year} {2012})}\BibitemShut {NoStop}%
\bibitem [{\citenamefont {Popova-Gorelova}\ \emph {et~al.}(2018)\citenamefont {Popova-Gorelova}, \citenamefont {Reis},\ and\ \citenamefont {Santra}}]{popova2018theory}%
  \BibitemOpen
  \bibfield  {author} {\bibinfo {author} {\bibfnamefont {D.}~\bibnamefont {Popova-Gorelova}}, \bibinfo {author} {\bibfnamefont {D.~A.}\ \bibnamefont {Reis}},\ and\ \bibinfo {author} {\bibfnamefont {R.}~\bibnamefont {Santra}},\ }\bibfield  {title} {\bibinfo {title} {Theory of x-ray scattering from laser-driven electronic systems},\ }\href@noop {} {\bibfield  {journal} {\bibinfo  {journal} {Physical Review B}\ }\textbf {\bibinfo {volume} {98}},\ \bibinfo {pages} {224302} (\bibinfo {year} {2018})}\BibitemShut {NoStop}%
\bibitem [{\citenamefont {Popova-Gorelova}\ and\ \citenamefont {Santra}(2024{\natexlab{a}})}]{Popova-GorelovaCommPhys24}%
  \BibitemOpen
  \bibfield  {author} {\bibinfo {author} {\bibfnamefont {D.}~\bibnamefont {Popova-Gorelova}}\ and\ \bibinfo {author} {\bibfnamefont {R.}~\bibnamefont {Santra}},\ }\bibfield  {title} {\bibinfo {title} {Atomic-scale imaging of laser-driven electron dynamics in solids},\ }\href {https://doi.org/10.1038/s42005-024-01810-7} {\bibfield  {journal} {\bibinfo  {journal} {Communications Physics}\ }\textbf {\bibinfo {volume} {7}},\ \bibinfo {pages} {317} (\bibinfo {year} {2024}{\natexlab{a}})}\BibitemShut {NoStop}%
\bibitem [{\citenamefont {Boemer}\ \emph {et~al.}(2021)\citenamefont {Boemer}, \citenamefont {Krebs}, \citenamefont {Benediktovitch}, \citenamefont {Rossi}, \citenamefont {Huotari},\ and\ \citenamefont {Rohringer}}]{Boemer}%
  \BibitemOpen
  \bibfield  {author} {\bibinfo {author} {\bibfnamefont {C.}~\bibnamefont {Boemer}}, \bibinfo {author} {\bibfnamefont {D.}~\bibnamefont {Krebs}}, \bibinfo {author} {\bibfnamefont {A.}~\bibnamefont {Benediktovitch}}, \bibinfo {author} {\bibfnamefont {E.}~\bibnamefont {Rossi}}, \bibinfo {author} {\bibfnamefont {S.}~\bibnamefont {Huotari}},\ and\ \bibinfo {author} {\bibfnamefont {N.}~\bibnamefont {Rohringer}},\ }\bibfield  {title} {\bibinfo {title} {Towards novel probes for valence charges via x-ray optical wave mixing},\ }\href {https://doi.org/10.1039/D0FD00130A} {\bibfield  {journal} {\bibinfo  {journal} {Faraday Discuss.}\ }\textbf {\bibinfo {volume} {228}},\ \bibinfo {pages} {451} (\bibinfo {year} {2021})}\BibitemShut {NoStop}%
\bibitem [{\citenamefont {Krebs}\ and\ \citenamefont {Rohringer}(2021)}]{krebs}%
  \BibitemOpen
  \bibfield  {author} {\bibinfo {author} {\bibfnamefont {D.}~\bibnamefont {Krebs}}\ and\ \bibinfo {author} {\bibfnamefont {N.}~\bibnamefont {Rohringer}},\ }\bibfield  {title} {\bibinfo {title} {Theory of parametric x-ray optical wavemixing processes},\ }\href@noop {} {\bibfield  {journal} {\bibinfo  {journal} {arXiv preprint arXiv:2104.05838}\ } (\bibinfo {year} {2021})}\BibitemShut {NoStop}%
\bibitem [{\citenamefont {Tamasaku}\ \emph {et~al.}(2011)\citenamefont {Tamasaku}, \citenamefont {Sawada}, \citenamefont {Nishibori},\ and\ \citenamefont {Ishikawa}}]{tamasaku2011visualizing}%
  \BibitemOpen
  \bibfield  {author} {\bibinfo {author} {\bibfnamefont {K.}~\bibnamefont {Tamasaku}}, \bibinfo {author} {\bibfnamefont {K.}~\bibnamefont {Sawada}}, \bibinfo {author} {\bibfnamefont {E.}~\bibnamefont {Nishibori}},\ and\ \bibinfo {author} {\bibfnamefont {T.}~\bibnamefont {Ishikawa}},\ }\bibfield  {title} {\bibinfo {title} {Visualizing the local optical response to extreme-ultraviolet radiation with a resolution of $\lambda$/380},\ }\href@noop {} {\bibfield  {journal} {\bibinfo  {journal} {Nature physics}\ }\textbf {\bibinfo {volume} {7}},\ \bibinfo {pages} {705} (\bibinfo {year} {2011})}\BibitemShut {NoStop}%
\bibitem [{\citenamefont {Schori}\ \emph {et~al.}(2017)\citenamefont {Schori}, \citenamefont {B\"omer}, \citenamefont {Borodin}, \citenamefont {Collins}, \citenamefont {Detlefs}, \citenamefont {Moretti~Sala}, \citenamefont {Yudovich},\ and\ \citenamefont {Shwartz}}]{SchoriXPDtoOpt}%
  \BibitemOpen
  \bibfield  {author} {\bibinfo {author} {\bibfnamefont {A.}~\bibnamefont {Schori}}, \bibinfo {author} {\bibfnamefont {C.}~\bibnamefont {B\"omer}}, \bibinfo {author} {\bibfnamefont {D.}~\bibnamefont {Borodin}}, \bibinfo {author} {\bibfnamefont {S.~P.}\ \bibnamefont {Collins}}, \bibinfo {author} {\bibfnamefont {B.}~\bibnamefont {Detlefs}}, \bibinfo {author} {\bibfnamefont {M.}~\bibnamefont {Moretti~Sala}}, \bibinfo {author} {\bibfnamefont {S.}~\bibnamefont {Yudovich}},\ and\ \bibinfo {author} {\bibfnamefont {S.}~\bibnamefont {Shwartz}},\ }\bibfield  {title} {\bibinfo {title} {Parametric down-conversion of x rays into the optical regime},\ }\href {https://doi.org/10.1103/PhysRevLett.119.253902} {\bibfield  {journal} {\bibinfo  {journal} {Phys. Rev. Lett.}\ }\textbf {\bibinfo {volume} {119}},\ \bibinfo {pages} {253902} (\bibinfo {year} {2017})}\BibitemShut {NoStop}%
\bibitem [{\citenamefont {Sofer}\ \emph {et~al.}(2019)\citenamefont {Sofer}, \citenamefont {Sefi}, \citenamefont {Strizhevsky}, \citenamefont {Aknin}, \citenamefont {Collins}, \citenamefont {Nisbet}, \citenamefont {Detlefs}, \citenamefont {Sahle},\ and\ \citenamefont {Shwartz}}]{SoferPDCUV}%
  \BibitemOpen
  \bibfield  {author} {\bibinfo {author} {\bibfnamefont {S.}~\bibnamefont {Sofer}}, \bibinfo {author} {\bibfnamefont {O.}~\bibnamefont {Sefi}}, \bibinfo {author} {\bibfnamefont {E.}~\bibnamefont {Strizhevsky}}, \bibinfo {author} {\bibfnamefont {H.}~\bibnamefont {Aknin}}, \bibinfo {author} {\bibfnamefont {S.~P.}\ \bibnamefont {Collins}}, \bibinfo {author} {\bibfnamefont {G.}~\bibnamefont {Nisbet}}, \bibinfo {author} {\bibfnamefont {B.}~\bibnamefont {Detlefs}}, \bibinfo {author} {\bibfnamefont {C.~J.}\ \bibnamefont {Sahle}},\ and\ \bibinfo {author} {\bibfnamefont {S.}~\bibnamefont {Shwartz}},\ }\bibfield  {title} {\bibinfo {title} {Observation of strong nonlinear interactions in parametric down-conversion of x-rays into ultraviolet radiation},\ }\href@noop {} {\bibfield  {journal} {\bibinfo  {journal} {Nature Communications}\ }\textbf {\bibinfo {volume} {10}},\ \bibinfo {pages} {5673} (\bibinfo {year} {2019})}\BibitemShut {NoStop}%
\bibitem [{\citenamefont {Jackson}(1998)}]{Jackson}%
  \BibitemOpen
  \bibfield  {author} {\bibinfo {author} {\bibfnamefont {J.~D.}\ \bibnamefont {Jackson}},\ }\href@noop {} {\emph {\bibinfo {title} {Classical electrodynamics}}}\ (\bibinfo  {publisher} {John Wiley \& Sons},\ \bibinfo {year} {1998})\BibitemShut {NoStop}%
\bibitem [{\citenamefont {Chollet}\ \emph {et~al.}(2015)\citenamefont {Chollet}, \citenamefont {Alonso-Mori}, \citenamefont {Cammarata}, \citenamefont {Damiani}, \citenamefont {Defever}, \citenamefont {Delor}, \citenamefont {Feng}, \citenamefont {Glownia}, \citenamefont {Langton}, \citenamefont {Nelson} \emph {et~al.}}]{XPP}%
  \BibitemOpen
  \bibfield  {author} {\bibinfo {author} {\bibfnamefont {M.}~\bibnamefont {Chollet}}, \bibinfo {author} {\bibfnamefont {R.}~\bibnamefont {Alonso-Mori}}, \bibinfo {author} {\bibfnamefont {M.}~\bibnamefont {Cammarata}}, \bibinfo {author} {\bibfnamefont {D.}~\bibnamefont {Damiani}}, \bibinfo {author} {\bibfnamefont {J.}~\bibnamefont {Defever}}, \bibinfo {author} {\bibfnamefont {J.~T.}\ \bibnamefont {Delor}}, \bibinfo {author} {\bibfnamefont {Y.}~\bibnamefont {Feng}}, \bibinfo {author} {\bibfnamefont {J.~M.}\ \bibnamefont {Glownia}}, \bibinfo {author} {\bibfnamefont {J.~B.}\ \bibnamefont {Langton}}, \bibinfo {author} {\bibfnamefont {S.}~\bibnamefont {Nelson}}, \emph {et~al.},\ }\bibfield  {title} {\bibinfo {title} {The x-ray pump--probe instrument at the linac coherent light source},\ }\href@noop {} {\bibfield  {journal} {\bibinfo  {journal} {Synchrotron Radiation}\ }\textbf {\bibinfo {volume} {22}},\ \bibinfo {pages} {503} (\bibinfo {year} {2015})}\BibitemShut {NoStop}%
\bibitem [{\citenamefont {Hartmann}\ \emph {et~al.}(2014)\citenamefont {Hartmann}, \citenamefont {Helml}, \citenamefont {Galler}, \citenamefont {Bionta}, \citenamefont {Gr{\"u}nert}, \citenamefont {L.~Molodtsov}, \citenamefont {Ferguson}, \citenamefont {Schorb}, \citenamefont {Swiggers}, \citenamefont {Carron} \emph {et~al.}}]{TimeTool}%
  \BibitemOpen
  \bibfield  {author} {\bibinfo {author} {\bibfnamefont {N.}~\bibnamefont {Hartmann}}, \bibinfo {author} {\bibfnamefont {W.}~\bibnamefont {Helml}}, \bibinfo {author} {\bibfnamefont {A.}~\bibnamefont {Galler}}, \bibinfo {author} {\bibfnamefont {M.~R.}\ \bibnamefont {Bionta}}, \bibinfo {author} {\bibfnamefont {J.}~\bibnamefont {Gr{\"u}nert}}, \bibinfo {author} {\bibfnamefont {S.}~\bibnamefont {L.~Molodtsov}}, \bibinfo {author} {\bibfnamefont {K.}~\bibnamefont {Ferguson}}, \bibinfo {author} {\bibfnamefont {S.}~\bibnamefont {Schorb}}, \bibinfo {author} {\bibfnamefont {M.}~\bibnamefont {Swiggers}}, \bibinfo {author} {\bibfnamefont {S.}~\bibnamefont {Carron}}, \emph {et~al.},\ }\bibfield  {title} {\bibinfo {title} {Sub-femtosecond precision measurement of relative x-ray arrival time for free-electron lasers},\ }\href@noop {} {\bibfield  {journal} {\bibinfo  {journal} {Nature photonics}\ }\textbf {\bibinfo {volume} {8}},\ \bibinfo {pages} {706} (\bibinfo {year} {2014})}\BibitemShut {NoStop}%
\bibitem [{\citenamefont {Momma}\ and\ \citenamefont {Izumi}(2011)}]{momma2011vesta}%
  \BibitemOpen
  \bibfield  {author} {\bibinfo {author} {\bibfnamefont {K.}~\bibnamefont {Momma}}\ and\ \bibinfo {author} {\bibfnamefont {F.}~\bibnamefont {Izumi}},\ }\bibfield  {title} {\bibinfo {title} {Vesta 3 for three-dimensional visualization of crystal, volumetric and morphology data},\ }\href@noop {} {\bibfield  {journal} {\bibinfo  {journal} {Journal of applied crystallography}\ }\textbf {\bibinfo {volume} {44}},\ \bibinfo {pages} {1272} (\bibinfo {year} {2011})}\BibitemShut {NoStop}%
\bibitem [{\citenamefont {You}\ \emph {et~al.}(2017)\citenamefont {You}, \citenamefont {Reis},\ and\ \citenamefont {Ghimire}}]{YouMgO}%
  \BibitemOpen
  \bibfield  {author} {\bibinfo {author} {\bibfnamefont {Y.~S.}\ \bibnamefont {You}}, \bibinfo {author} {\bibfnamefont {D.~A.}\ \bibnamefont {Reis}},\ and\ \bibinfo {author} {\bibfnamefont {S.}~\bibnamefont {Ghimire}},\ }\bibfield  {title} {\bibinfo {title} {Anisotropic high-harmonic generation in bulk crystals},\ }\href@noop {} {\bibfield  {journal} {\bibinfo  {journal} {Nature physics}\ }\textbf {\bibinfo {volume} {13}},\ \bibinfo {pages} {345} (\bibinfo {year} {2017})}\BibitemShut {NoStop}%
\bibitem [{\citenamefont {Ghimire}\ and\ \citenamefont {Reis}(2019)}]{GhimireAndReisReview}%
  \BibitemOpen
  \bibfield  {author} {\bibinfo {author} {\bibfnamefont {S.}~\bibnamefont {Ghimire}}\ and\ \bibinfo {author} {\bibfnamefont {D.~A.}\ \bibnamefont {Reis}},\ }\bibfield  {title} {\bibinfo {title} {High-harmonic generation from solids},\ }\href@noop {} {\bibfield  {journal} {\bibinfo  {journal} {Nature physics}\ }\textbf {\bibinfo {volume} {15}},\ \bibinfo {pages} {10} (\bibinfo {year} {2019})}\BibitemShut {NoStop}%
\bibitem [{\citenamefont {Basov}\ \emph {et~al.}(2017)\citenamefont {Basov}, \citenamefont {Averitt},\ and\ \citenamefont {Hsieh}}]{Basov}%
  \BibitemOpen
  \bibfield  {author} {\bibinfo {author} {\bibfnamefont {D.}~\bibnamefont {Basov}}, \bibinfo {author} {\bibfnamefont {R.}~\bibnamefont {Averitt}},\ and\ \bibinfo {author} {\bibfnamefont {D.}~\bibnamefont {Hsieh}},\ }\bibfield  {title} {\bibinfo {title} {Towards properties on demand in quantum materials},\ }\href@noop {} {\bibfield  {journal} {\bibinfo  {journal} {Nature materials}\ }\textbf {\bibinfo {volume} {16}},\ \bibinfo {pages} {1077} (\bibinfo {year} {2017})}\BibitemShut {NoStop}%
\bibitem [{\citenamefont {Huang}\ \emph {et~al.}(2017)\citenamefont {Huang}, \citenamefont {Ding}, \citenamefont {Feng}, \citenamefont {Hemsing}, \citenamefont {Huang}, \citenamefont {Krzywinski}, \citenamefont {Lutman}, \citenamefont {Marinelli}, \citenamefont {Maxwell},\ and\ \citenamefont {Zhu}}]{LCLS1stHardAtto}%
  \BibitemOpen
  \bibfield  {author} {\bibinfo {author} {\bibfnamefont {S.}~\bibnamefont {Huang}}, \bibinfo {author} {\bibfnamefont {Y.}~\bibnamefont {Ding}}, \bibinfo {author} {\bibfnamefont {Y.}~\bibnamefont {Feng}}, \bibinfo {author} {\bibfnamefont {E.}~\bibnamefont {Hemsing}}, \bibinfo {author} {\bibfnamefont {Z.}~\bibnamefont {Huang}}, \bibinfo {author} {\bibfnamefont {J.}~\bibnamefont {Krzywinski}}, \bibinfo {author} {\bibfnamefont {A.}~\bibnamefont {Lutman}}, \bibinfo {author} {\bibfnamefont {A.}~\bibnamefont {Marinelli}}, \bibinfo {author} {\bibfnamefont {T.}~\bibnamefont {Maxwell}},\ and\ \bibinfo {author} {\bibfnamefont {D.}~\bibnamefont {Zhu}},\ }\bibfield  {title} {\bibinfo {title} {Generating single-spike hard x-ray pulses with nonlinear bunch compression in free-electron lasers},\ }\href@noop {} {\bibfield  {journal} {\bibinfo  {journal} {Physical Review Letters}\ }\textbf {\bibinfo {volume} {119}},\ \bibinfo {pages} {154801} (\bibinfo {year} {2017})}\BibitemShut {NoStop}%
\bibitem [{\citenamefont {Yan}\ \emph {et~al.}(2024)\citenamefont {Yan}, \citenamefont {Qin}, \citenamefont {Chen}, \citenamefont {Decking}, \citenamefont {Dijkstal}, \citenamefont {Guetg}, \citenamefont {Inoue}, \citenamefont {Kujala}, \citenamefont {Liu}, \citenamefont {Long}, \citenamefont {Mirian},\ and\ \citenamefont {Geloni}}]{EUXFELAtto}%
  \BibitemOpen
  \bibfield  {author} {\bibinfo {author} {\bibfnamefont {J.}~\bibnamefont {Yan}}, \bibinfo {author} {\bibfnamefont {W.}~\bibnamefont {Qin}}, \bibinfo {author} {\bibfnamefont {Y.}~\bibnamefont {Chen}}, \bibinfo {author} {\bibfnamefont {W.}~\bibnamefont {Decking}}, \bibinfo {author} {\bibfnamefont {P.}~\bibnamefont {Dijkstal}}, \bibinfo {author} {\bibfnamefont {M.}~\bibnamefont {Guetg}}, \bibinfo {author} {\bibfnamefont {I.}~\bibnamefont {Inoue}}, \bibinfo {author} {\bibfnamefont {N.}~\bibnamefont {Kujala}}, \bibinfo {author} {\bibfnamefont {S.}~\bibnamefont {Liu}}, \bibinfo {author} {\bibfnamefont {T.}~\bibnamefont {Long}}, \bibinfo {author} {\bibfnamefont {N.}~\bibnamefont {Mirian}},\ and\ \bibinfo {author} {\bibfnamefont {G.}~\bibnamefont {Geloni}},\ }\bibfield  {title} {\bibinfo {title} {Terawatt-attosecond hard x-ray free-electron laser at high repetition rate},\ }\href {https://doi.org/10.1038/s41566-024-01566-0} {\bibfield  {journal} {\bibinfo  {journal} {Nature Photonics}\ }\textbf {\bibinfo {volume}
  {18}},\ \bibinfo {pages} {1293} (\bibinfo {year} {2024})}\BibitemShut {NoStop}%
\bibitem [{\citenamefont {Zhu}\ and\ \citenamefont {Reis}(2024)}]{attovision}%
  \BibitemOpen
  \bibfield  {author} {\bibinfo {author} {\bibfnamefont {D.}~\bibnamefont {Zhu}}\ and\ \bibinfo {author} {\bibfnamefont {D.~A.}\ \bibnamefont {Reis}},\ }\bibfield  {title} {\bibinfo {title} {Attosecond x-ray laser vision},\ }\href {https://doi.org/10.1038/s41566-024-01575-z} {\bibfield  {journal} {\bibinfo  {journal} {Nature Photonics}\ }\textbf {\bibinfo {volume} {18}},\ \bibinfo {pages} {1232} (\bibinfo {year} {2024})}\BibitemShut {NoStop}%
\bibitem [{\citenamefont {Popova-Gorelova}\ and\ \citenamefont {Santra}(2024{\natexlab{b}})}]{popova2024microscopic}%
  \BibitemOpen
  \bibfield  {author} {\bibinfo {author} {\bibfnamefont {D.}~\bibnamefont {Popova-Gorelova}}\ and\ \bibinfo {author} {\bibfnamefont {R.}~\bibnamefont {Santra}},\ }\bibfield  {title} {\bibinfo {title} {Microscopic nonlinear optical response: Analysis and calculations with the floquet--bloch formalism},\ }\href@noop {} {\bibfield  {journal} {\bibinfo  {journal} {Structural Dynamics}\ }\textbf {\bibinfo {volume} {11}} (\bibinfo {year} {2024}{\natexlab{b}})}\BibitemShut {NoStop}%
\bibitem [{\citenamefont {Levine}\ and\ \citenamefont {Allan}(1989)}]{levine1989linear}%
  \BibitemOpen
  \bibfield  {author} {\bibinfo {author} {\bibfnamefont {Z.~H.}\ \bibnamefont {Levine}}\ and\ \bibinfo {author} {\bibfnamefont {D.~C.}\ \bibnamefont {Allan}},\ }\bibfield  {title} {\bibinfo {title} {Linear optical response in silicon and germanium including self-energy effects},\ }\href@noop {} {\bibfield  {journal} {\bibinfo  {journal} {Physical review letters}\ }\textbf {\bibinfo {volume} {63}},\ \bibinfo {pages} {1719} (\bibinfo {year} {1989})}\BibitemShut {NoStop}%
\bibitem [{\citenamefont {Phillips}(1962)}]{phillips1962band}%
  \BibitemOpen
  \bibfield  {author} {\bibinfo {author} {\bibfnamefont {J.}~\bibnamefont {Phillips}},\ }\bibfield  {title} {\bibinfo {title} {Band structure of silicon, germanium, and related semiconductors},\ }\href@noop {} {\bibfield  {journal} {\bibinfo  {journal} {Physical Review}\ }\textbf {\bibinfo {volume} {125}},\ \bibinfo {pages} {1931} (\bibinfo {year} {1962})}\BibitemShut {NoStop}%
\bibitem [{\citenamefont {Hybertsen}\ and\ \citenamefont {Louie}(1985)}]{hybertsen1985first}%
  \BibitemOpen
  \bibfield  {author} {\bibinfo {author} {\bibfnamefont {M.~S.}\ \bibnamefont {Hybertsen}}\ and\ \bibinfo {author} {\bibfnamefont {S.~G.}\ \bibnamefont {Louie}},\ }\bibfield  {title} {\bibinfo {title} {First-principles theory of quasiparticles: calculation of band gaps in semiconductors and insulators},\ }\href@noop {} {\bibfield  {journal} {\bibinfo  {journal} {Physical review letters}\ }\textbf {\bibinfo {volume} {55}},\ \bibinfo {pages} {1418} (\bibinfo {year} {1985})}\BibitemShut {NoStop}%
\bibitem [{\citenamefont {Warren}(1990)}]{warren1990x}%
  \BibitemOpen
  \bibfield  {author} {\bibinfo {author} {\bibfnamefont {B.~E.}\ \bibnamefont {Warren}},\ }\href@noop {} {\emph {\bibinfo {title} {X-ray Diffraction}}}\ (\bibinfo  {publisher} {Courier Corporation},\ \bibinfo {year} {1990})\BibitemShut {NoStop}%
\end{thebibliography}%

\end{document}